\documentclass[longauth]{aa}  
\usepackage{booktabs} 
\usepackage{graphicx}

\usepackage[normalem]{ulem}
\usepackage{xcolor}

\usepackage{txfonts}
\usepackage{natbib}
\bibpunct{(}{)}{;}{a}{}{,}
\usepackage[breaklinks, colorlinks, citecolor=blue, linkcolor=blue]{hyperref}
\usepackage{threeparttable}
\usepackage{xcolor}
\usepackage{orcidlink}
\usepackage{siunitx}
\usepackage{url}

\begin{document} 

   \title{Tighter constraints on the atmosphere of GJ\,436\,b from combined high-resolution CARMENES and CRIRES$^+$ observations}\titlerunning{Characterization of the atmosphere of GJ\,436\,b}\authorrunning{A. Peláez-Torres et al.}

   \author{A.\,Peláez-Torres
          \inst{\ref{IAA_CSIC}}\thanks{\texttt{E-mail: \color{magenta}apelaez@iaa.es}}\,\orcidlink{0000-0001-9204-8498},
          A.\,Sánchez-López\inst{\ref{IAA_CSIC}}\,\orcidlink{0000-0002-0516-7956},
          L.\,Nortmann\inst{\ref{IAG}},
          M.\,López-Puertas\inst{\ref{IAA_CSIC}},
          E.\,González-Álvarez\inst{\ref{CAB}}\,\orcidlink{0000-0002-4820-2053},
          H.\,M.\,Tabernero\inst{\ref{IEEC},\ref{UAB}},
          C.\,Jiang\inst{\ref{IAC},\ref{ULL}},
          D.\,Revilla\inst{\ref{IAA_CSIC}},
          G.\,Morello\inst{\ref{IAA_CSIC},\ref{INAF-Palermo}},
          J.\,Orell-Miquel\inst{\ref{Austin}},
          E.\,Pallé\inst{\ref{IAC},\ref{ULL}},
          P.\,J.\,Amado\inst{\ref{IAA_CSIC}},
          J.\,A.\,Caballero\inst{\ref{CAB}},
          I.\,Ribas\inst{\ref{UAB},\ref{IEEC}},
          A.\,Reiners\inst{\ref{IAG}},
          A.\,Quirrenbach\inst{\ref{Heidelberg}},
          D.\,Cont\inst{\ref{Maximilians},\ref{Garching}},
          S. Dreizler\inst{\ref{IAG}},
          A.\,Fernández-Martín\inst{\ref{CAHA}},
          A.\,P.\,Hatzes\inst{\ref{OKS}},
          Th.\,Henning\inst{\ref{MPI}},
          F.\,Lesjak\inst{\ref{AIP},\ref{IAG}},
          D.\,Montes\inst{\ref{UCM}},
          A.\,Schweizer\inst{\ref{Hamburg}}, 
          T.\,Trifonov\inst{\ref{Heidelberg}, \ref{bulgaria}},
          and
          F.\,Yan\inst{\ref{China}}          
          }

    \institute{Instituto de Astrofísica de Andalucía (IAA-CSIC), Glorieta de la Astronomía s/n, Genil, E-18008 Granada, Spain \label{IAA_CSIC}
    \and Institut für Astrophysik und Geophysik, Georg-August-Universität, Friedrich-Hund-Platz 1, 37077 Göttingen, Germany \label{IAG}
    \and Centro de Astrobiolog\'ia (CSIC-INTA), ESAC, Camino bajo del castillo s/n, 28692 Villanueva de la Cañada, Madrid, Spain \label{CAB}
    \and Institut d’Estudis Espacials de Catalunya (IEEC), 08034 Barcelona, Spain \label{IEEC}
    \and Institut de Ci\`encies de l’Espai (CSIC-IEEC), Campus UAB, c/ de Can Magrans s/n, 08193 Bellaterra, Barcelona, Spain \label{UAB}
    \and Instituto de Astrof\'isica de Canarias (IAC), 38200 La Laguna, Tenerife, Spain \label{IAC}
    \and Departamento de Astrof\'isica, Universidad de La Laguna (ULL), 38206 La Laguna, Tenerife, Spain \label{ULL}
    \and INAF- Palermo Astronomical Observatory, Piazza del Parlamento 1, 90134 Palermo, Italy \label{INAF-Palermo}
    \and Department of Astronomy, University of Texas at Austin, 2515 Speedway, Austin, TX 78712, USA \label{Austin}
    \and Landessternwarte, Zentrum für Astronomie der Universität Heidelberg, Königstuhl 12,
    69117 Heidelberg, Germany \label{Heidelberg}
    \and Universitäts-Sternwarte, Fakultät für Physik, Ludwig-Maximilians-Universität München, Scheinerstr 1, 81679 München, Germany \label{Maximilians}
    \and Exzellenzcluster Origins, Boltzmannstrasse 2, 85748 Garching, Germany \label{Garching}
    \and Centro Astronómico Hispano en Andalucía (CAHA), Observatorio de CalarAlto, Sierra de los Filabres, 04550 Gérgal, Spain \label{CAHA}
    \and Thüringer Landessternwarte Tautenburg, Sternwarte 5, 07778 Tautenburg, Germany \label{OKS}
    \and Max-Planck-Institut für Astronomie, Königstuhl 17, 69117 Heidelberg, Germany \label{MPI}
    \and Leibniz Institute for Astrophysics Potsdam (AIP), An der Sternwarte 16, 14482 Potsdam, Germany \label{AIP}
    \and Departamento de F\'{i}sica de la Tierra y Astrof\'{i}sica 
    and IPARCOS-UCM (Instituto de F\'{i}sica de Part\'{i}culas y del Cosmos de la UCM), Facultad de Ciencias F\'{i}sicas, Universidad Complutense de Madrid, 28040, Madrid, Spain \label{UCM}
    \and Hamburger Sternwarte, Gojenbergsweg 112, 21029 Hamburg, Germany \label{Hamburg}
    \and Department of Astronomy, Faculty of Physics, Sofia University “St Kliment Ohridski”, 5 James Bourchier Blvd., 1164 Sofia, Bulgaria \label{bulgaria}
    \and Department of Astronomy, University of Science and Technology of China, Hefei 230026, PR China \label{China}
             }

   \date{Received 06 October 2025}
 
  \abstract
   {Transmission spectra of Neptune-sized exoplanets are frequently observed to be featureless at low-to-mid resolutions from space; whereas high-altitude clouds can mute spectral features, high atmospheric metallicities can also result in compressed envelopes, where low scale heights may also yield undetectable signatures.}
   {We aim to study the atmospheric properties of the warm Neptune GJ\,436\,b by combining a set of five transit events observed with the CARMENES spectrograph with one transit from CRIRES$^+$ so as to provide the most constrained results possible at high resolution.}
   {We removed telluric and stellar signals from the data using {\tt SysRem} and potential planetary signals were investigated using the cross-correlation technique. Following standard procedures for undetected species, we performed injection recovery tests and Bayesian retrievals to place constraints on the detectability of the main near-infrared absorbers. In addition, we simulated ELT/ANDES observations by computing end-to-end in silico datasets with {\tt EXoPLORE}.}
   {No molecular signals were detected in the atmosphere of GJ\,436\,b, which is consistent with previous studies. Combined CARMENES–CRIRES$^+$ injection-recovery and Bayesian retrieval analyses show that the atmosphere is likely covered by high-altitude clouds ($\sim$\,$1$\,mbar) at low and intermediate metallicities or, alternatively, is very metal-rich ($\gtrsim$\,$900\times$ solar), which would suppress spectral features without invoking clouds. Simulations of ELT/ANDES observations suggest a boost by nearly an order of magnitude to the upper limit in the photon-limited regime, reaching $0.1$\,mbar at $10$--$300\times$\,solar metallicities.}
   {The joint analysis of all useful transit observations from CARMENES and CRIRES$^+$ provides the most stringent constraints to date on the atmospheric properties of GJ\,436\,b. Complementary CCF-based and retrieval approaches consistently indicate that the atmosphere is either cloudy or highly metal enriched. Any weak near-infrared absorption lines, if present, are likely to be below current detection limits. However, according to our simulations, these features may be revealed with ELT/ANDES even in single-transit observations.}

   \keywords{planets and satellites: atmospheres --
                planets and satellites: individual: GJ\,436\,b --
                techniques: spectroscopic -- methods: observational -- methods: statistical             
               }

   \maketitle

\section{Introduction}

Planets similar to and smaller than Neptune present a great challenge for atmospheric characterization due to their generally low transit signals, when compared to the widely studied hot and ultra-hot Jupiters \citep{bean2010, berta2012, kreidberg2014, benneke2019a, benneke2024, gandhi2020, madhusudhan2021, madhusudhan2023, dash2024, grasser2024}.
Following the discovery of GJ\,1214\,b \citep[$8.14$\,M$_\oplus$;][]{charbonneau2009}, a benchmark sub-Neptune orbiting an M dwarf, a thorough analysis of its atmosphere using the \textit{Hubble} Space Telescope (HST) revealed a featureless transmission spectrum \citep{kreidberg2014}. Since then, multiple Neptune- and sub-Neptune-like planets have shown similar properties \citep{demory2016, guilluy2021, alam2022, kreidberg2022, brande2022}. This lack of spectroscopic features is often explained by the presence of high-altitude cloud decks, highly compressed (i.e. high-mean-molecular-weight) atmospheres, or even by these exoplanets being airless bodies \citep{knutson2014a, knutson2014b, wakeford2017, benneke2019a, kreidberg2022}. Indeed, rocky planets with significant atmospheres and gas giants in the Solar System generally have widespread cloud coverage \citep{irwin2009, marley2013}.

However, observations with the \textit{James Webb} Space Telescope \citep[JWST;][]{gardner2006} using the NIRCam instrument have revealed the presence of water vapour (H$_2$O), methane (CH$_4$), sulphur dioxide (SO$_2$), and carbon monoxide (CO) in the atmosphere of the Neptune-like planet GJ\,3470\,b, which has an estimated atmospheric metallicity of approximately $100\times$\,solar. This contrasts with prior HST and \textit{Spitzer} observations, which suggested an atmospheric metallicity close to solar \citep{benneke2019a, beatty2024}. Similarly, highly metal-enriched envelopes have been inferred for several sub-Neptunes, including TOI--270\,d \citep[$50$\% metals, $\sim$\,$500\times$\,solar;][]{benneke2024}, GJ\,1214\,b \citep[$\sim$\,$1000\times$\,solar;][]{kempton2023, schlawin2024, ohno2025}, and K2--18\,b \citep[$\sim$\,$100$--$300\times$\,solar;][]{charnay2021}. In the case of K2--18\,b, an initial interpretation of its HST Wide Field Camera $3$ (WFC3) transmission spectrum suggested the presence of H$_2$O, but this signal was later determined to be CH$_4$ with JWST \citep{benneke2019b, barclay2021, madhusudhan2023}, whose absorption bands are difficult to distinguish from those of water vapour at the resolving power of WFC3. Besides CH$_4$, these recent JWST observations have led to the first confirmed detection of CO$_2$ in K2--18\,b \citep{benneke2019b, barclay2021, madhusudhan2023}.

Another interesting target in the Neptune-like range is GJ\,436\,b \citep{butler2004}. Comprehensive studies of the GJ\,436 planetary system have revealed its unusual properties. Namely, this transiting warm Neptune possesses a steeply polar orbit \citep{bourrier2022}, as well as a long hydrogen tail, observed in Ly$\alpha$, which results from its escaping atmosphere \citep{bourrier2016}. Spectroscopic observations have yielded non-detections of the metastable helium triplet He~{\sc i} $\lambda$1083\,nm \citep{nortmann2018, guilluy2024, masson2024}, which is possibly caused by a low He concentration \citep{rumenskikh2023}, and prevents further characterization of the escaping gas. Interestingly, a recent study by \cite{revilla2025} seems to provide evidence of  star-planet magnetic interactions, which lead to a constraint on the planet's magnetic field and magnetospheric radius \citep{loyd2023, vidotto2023}. Contrary to popular belief, the presence of a planetary magnetic field does not necessarily protect a planet from stellar wind erosion but it can even increase the rate of ion escape \citep{ramstad2021}.

The atmosphere of GJ\,436\,b was also the subject of early investigations with {\it Spitzer} transit and eclipse observations, which led to contrasting claims about the presence of CO and CH$_4$, as well as the influence of stellar variability on those measurements \citep{beaulieu2011, knutson2011}. Based on chemical-equilibrium modelling, \cite{stevenson2010} predicted that the atmosphere of GJ\,436\,b could be rich in CO, with a significant depletion of CH$_4$. This is somewhat unusual as CH$_4$ is expected to be the dominant carbon-bearing species in warm hydrogen-rich atmospheres \citep{seager2010}. Vertical mixing, combined with a high atmospheric metallicity (>$10\times$\,solar), could lead to enhanced CO abundances, while a depletion of CH$_4$ may arise via non-equilibrium chemical processes \citep{madhusudhan2011}. According to the model of \cite{moses2013}, the claimed composition could be consistent with a high metallicity in the range of $\sim$\,$230$--$2000\times$\,solar. High metallicities lead to a low atmospheric scale height ($H \propto m^{-1}$, where $m$ is the mean molecular mass), which reduces the amplitude of spectral features. However, reanalyses of the {\it Spitzer} data indicate a constant transit depth across both time and wavelength ($3$--\SI{24}{\micro\meter}; \citealt{lanotte2014, morello2015}). Furthermore, optical and near-infrared transmission spectra obtained with HST did not reveal any significant absorption features \citep{knutson2014a, lothringer2018}. More recently, GJ\,436\,b's atmosphere was analysed at low resolution spectroscopy (LRS) with JWST using Near Infrared Camera (NIRCam) and Mid Infrared Instrument (MIRI) by \cite{mukherjee2025}, revealing a tentative detection of CO$_2$ ($2\sigma$). However, the absence of strong transmission features in previous high-resolution spectroscopy (HRS) studies suggests the presence of clouds. These findings have led to a general consensus that GJ\,436\,b's atmosphere could contain a high-altitude cloud deck that mutes most of the spectral features across the near- (NIR) and mid-infrared (MIR) wavelength ranges.
    
High resolution spectroscopy has emerged as one of the most effective techniques for characterizing exoplanet atmospheres, both in transit and dayside geometries, across optical and infrared wavelengths \citep[e.g.][]{snellen2025exoplanet}. Furthermore, HRS is highly sensitive to small Doppler shifts in the spectral lines and to their shape, which enables us to gain information about exo-atmospheric circulation \citep{2010Natur.465.1049S, 2012ApJ...751..117M, 2019AJ....157..209F}. HRS characterization of exoplanet atmospheres has enabled the detection of single species \citep{charbonneau2002, snellen2008, snellen2010, brogi2012, birkby2013, de2013, lockwood2014, casasayas2021, stangret2022} and multiple molecules in the same atmosphere \citep{hawker2018, cabot2019, kesseli2020, giacobbe2021, cont2021, sanchez2022}, and has eventually evolved into the development of Bayesian retrievals \citep{brogi2017, brogi2019, gibson2020, gibson2022, cont2022, maguire2024, blain2024}. These retrieval techniques are capable of constraining relevant parameters (e.g. atmospheric elemental abundances, temperature structure and winds, and the planet's rotational velocity, among others) using Bayesian statistics. Their application has yielded robust statistical significance in detecting molecular signatures from close-orbiting gas giants observed by transmission and emission spectroscopy \citep{lesjak2023, cont2024, nortmann2025}. Furthermore, low-resolution retrievals \citep{madhusudhan2009, madhusudhan2014, evans2016, pinhas2018, wakeford2017, chachan2019, chubb2020} or photometric data \citep{cont2024, cont2025} can be combined with high-resolution studies to provide tighter constraints on the retrieved parameters \citep{brogi2017, boucher2023, smith2024}. In contrast to low-resolution studies, HRS data allow us to distinguish the core and wings of spectral lines, which makes this technique ideal to attempt detections in hazy exo-atmospheres where only line cores may protrude above the continua \citep[e.g.][]{pino2018a, pino2018b, sanchez2020}.
    
A recent study by \cite{grasser2024} using the high-resolution spectrograph CRIRES$^+$, on European Southern Observatory's (ESO) Very Large Telescope \citep{Kaeufl2004, follert2014, dorn2014, dorn2023} did not detect transmission signals of H$_2$O, CH$_4$, or CO in the atmosphere of GJ\,436\,b in the $1431$--$1837$\,nm spectral range covered by the H1567 setting. However, they were able to place upper limits on the atmospheric composition by performing injection recovery tests of model absorption signals, which reveals that CRIRES$^+$ observations should have detected water vapour if GJ\,436\,b presented a cloud deck at pressures higher than $10$\,mbar (lower altitudes) and if, simultaneously, the atmospheric metallicity was in the range of $1$--$300\times$\,solar.

In this paper, we compile all available datasets observed with the Calar Alto high-Resolution search for M dwarfs with Exo-earths with Near-infrared and optical Échelle Spectrographs \citep[CARMENES;][]{quirrenbach2014, quirrenbach2018} and all observed with the CRyogenic high-resolution InfraRed Echelle Spectrograph \citep[CRIRES$^+$;][]{Kaeufl2004, follert2014, dorn2014} for GJ\,436\,b (six transits in total) to explore the atmospheric composition of this exoplanet. To this end, we used the cross-correlation technique over multiple transit events.

The paper is structured as follows. In Sect.\,\ref{sec:observations}, we describe the observations. In Sect.\,\ref{sec:system_parameters}, we refine both the stellar and planetary parameters of the GJ\,436 system. The methods used for the analysis of the spectral data are detailed in Sect.\,\ref{sec:methods}, including those used for identifying the molecular signatures. We present and discuss our results in Sect.\,\ref{sec:results_discussion}, including future prospects for atmospheric characterization of cloudy sub-Neptunes with the ArmazoNes high Dispersion Echelle Spectrograph (ANDES) \citep{marconi2022andes, 2025ExA....59...29P} on the Extremely Large Telescope (ELT), and we outline our main conclusions are given in Sect.\,\ref{sec:conclusion}.

    \begin{figure*}[]
    \centering
    \includegraphics[width=\textwidth]{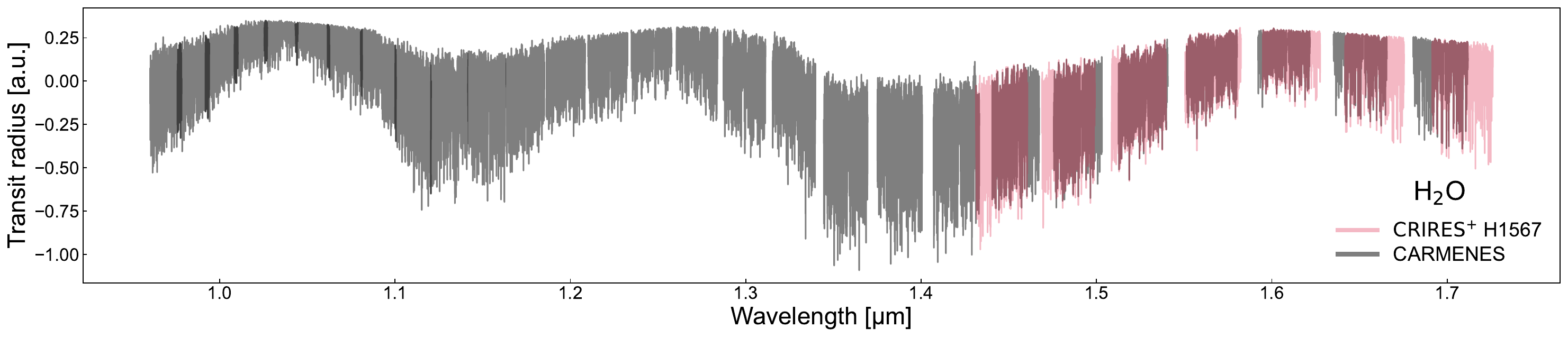}
    \caption{Spectral coverage of water vapour transmission models for GJ\,436\,b of the CARMENES NIR channel (grey) and the $\mathrm{CRIRES}^{+}$ H1567 setting (red). Similar figures for CH$_4$ and CO can be found in Appendix\,\ref{fig:range_comparison_ch4_co}.}
    \label{fig:range_comparison}
    \end{figure*}

\section{Observations}
\label{sec:observations}

\subsection{CARMENES}

\begin{figure}[]
    \centering
    \includegraphics[width=\columnwidth]{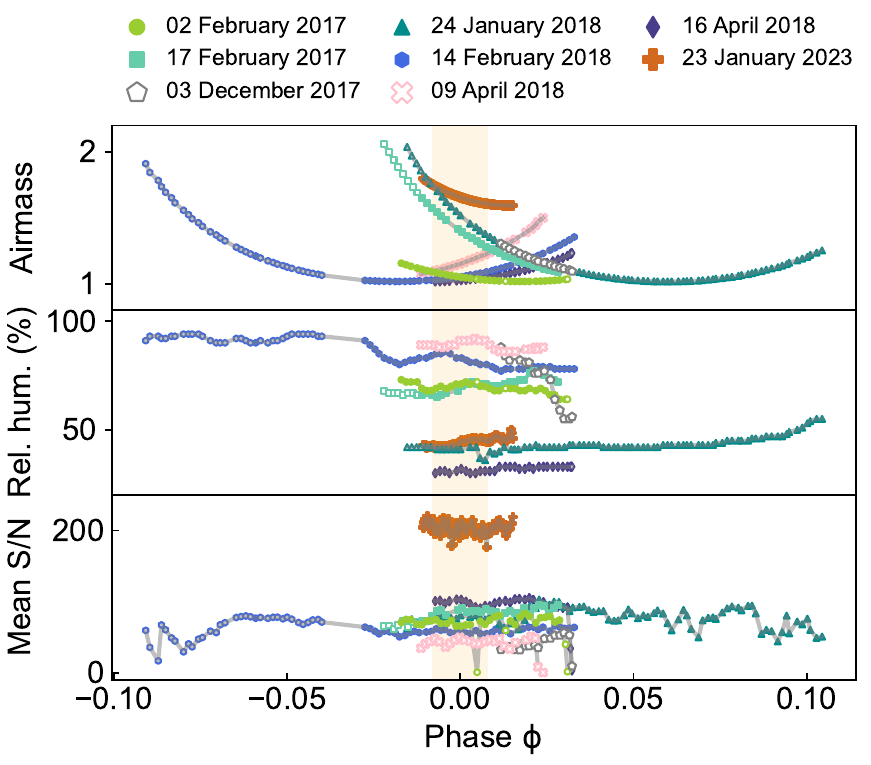}
    \caption{Evolution of the airmass, relative humidity, and mean S/N per spectrum over the entire CARMENES NIR range. The orange shaded area marks the in-transit times. Excluded spectra are represented with open symbols.}
    \label{fig:night_conditions}
\end{figure}

\begin{figure}[]
    \centering
    \includegraphics[width=\columnwidth]{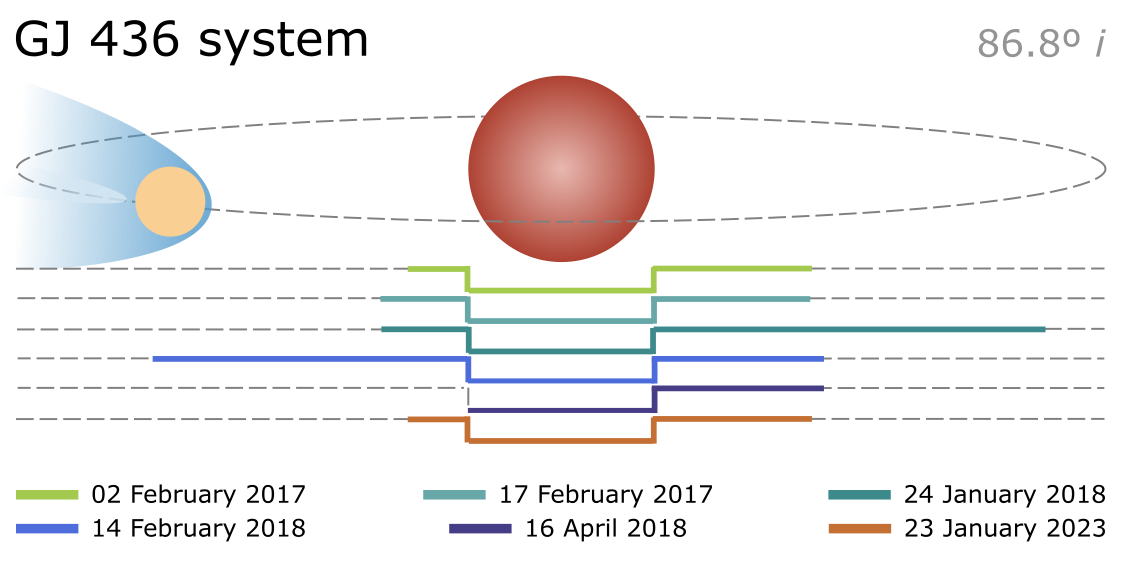}
    \caption{Schematic of transmission spectroscopy observations of GJ\,436 system used in this work. The orbital phase coverage for the different transits is shown in the lower part.
    }
    \label{fig:transits}
\end{figure}

We analysed five transit datasets of GJ\,436\,b observed with the CARMENES instrument, installed at the $3.5$\,m telescope of the Centro Astronómico Hispano en Andalucía (CAHA). CARMENES has two spectral channels: the optical channel (VIS), covering the wavelength range of $0.52$--\SI{0.96}{\micro\meter} in $55$ spectral orders ($\mathcal{R}$\,$\approx$\,$94\,600$), and the near-infrared channel (NIR), covering from $0.96$--\SI{1.71}{\micro\meter} in $28$ orders ($\mathcal{R}$\,$\approx$\,$80\,400$). The CARMENES NIR spectral coverage includes that of the CRIRES$^+$'s H1567 setting, as shown in Fig.\,\ref{fig:range_comparison}. Each CARMENES channel has two fibres: fibre A was used to observe the target star, while fibre B was placed on the sky (88\,arcsec to the west) to identify sky emission lines. Here, we focus on the observations conducted using the NIR channel, since this is the spectral region where the ro-vibrational lines from the spectroscopically active molecules are located. All of our datasets were reduced by the CARMENES pipeline, \texttt{caracal v$2.20$} \citep{caballero2016}. In Sect.\,\ref{sec:system_parameters}, we revised all stellar and planetary parameters required for this study. The five datasets used for GJ\,436\,b (see Table\,\ref{tab:observations} and Fig.\,\ref{fig:night_conditions}) covered the pre-, peri-, and post-transit phases of GJ\,436\,b, except for the pre-transit of the fifth night (see Fig.\,\ref{fig:transits}). A sixth ($03$ December $2017$) and a seventh ($09$ April $2018$) night were available, but Due to the lack of in-transit spectra from the sixth night, and the high relative humidity and low signal to noise ratio (S/N) of the seventh, we excluded them from our analysis. The S/N across all five used nights ranges from $\sim$\,$60$ to $\sim$\,$100$. However, during certain exposures on the first and last nights, it fell below $60$. Consequently, we excluded four exposures from the first night and three from the fifth to avoid low-quality spectra. During the course of all five nights, the star moved from airmass $1.02$ to $2.06$. We also excluded eight exposures from the second night and four from the third, as they were taken at an airmass greater than $1.75$. In addition, our fourth night showed a $30$ minutes gap (from $\phi=-0.040$ to $\phi=-0.027$) due to a guiding issue. Additionally, the S/N and relative humidity in the first part of that night exhibited irregularities. To ensure consistency, we also excluded that part of the night from our analysis.

\subsection{CRIRES$^+$}

We analysed an additional transit dataset observed with the CRIRES$^+$ spectrograph, installed at the ESO's Very Large Telescope (VLT) at Cerro Paranal. The observation was carried out on $23$ January $2023$ with the H1567 wavelength setting and it was previously analysed by \citet{grasser2024}. This setting covers the $1.49$\,--\,$1.78$\,$\mu$m range over eight spectral orders, at a resolving power of $\mathcal{R}$\,$\approx100\,000$. Further details on the observing conditions of this night are provided in Table\,\ref{tab:observations}. Two additional transit events were observed with CRIRES$^+$ but, as described by \citet{grasser2024}, their signal-to-noise ratio is rather low due to issues in the adaptive optics system. 

The CRIRES$^+$ raw data for GJ\,436\,b were retrieved from the ESO Archive\footnote{\url{https://archive.eso.org/cms/eso-data.html}} and reduced using the standard \texttt{CR2RES} pipeline routines (version 1.4.4) through the ESO Recipe Execution Tool (\texttt{EsoRex}, version 3.13.8). This pipeline includes dark calibration (not used during nodding subtraction), bad-pixel masking, flat calibration, wavelength calibration using a combination of the Fabry-Pérot etalon and uranium-neon lamp, and 1D spectral extraction.
The observation consisted of 7.5 nodding cycles, following the ``$\rm A_{1,2,3}$--$\rm B_{1,2,3}$--$\rm B_{4,5,6}$--$\rm A_{4,5,6}$'' nodding pattern. In order to perform nodding subtraction, we regrouped the nodding pairs as $\rm A_1B_1$, $\rm A_2B_2$, ... $\rm A_6B_6$. The spectra were extracted using the optimal extraction algorithm, with a fixed extraction height of 20 pixels ($\sim$8 times the median full width at half maximum of the point spread function along the slit) and no-light rows subtracted. Also, we discarded the bluest order \#9 (1.43--1.46 $\mu$m) due to incomplete wavelength coverage and strong telluric absorption. 
The observation was conducted in in adaptive-optics mode mode and that, due to seeing conditions ($\sim$0.65$''$), the slit was not evenly illuminated with light in the dispersion direction during the observation. Although this might cause a slight spectral shift between A and B positions (as described in the \texttt{CR2RES} user manual\footnote{\url{https://ftp.eso.org/pub/dfs/pipelines/instruments/cr2res/cr2re-pipeline-manual-1.6.10.pdf}}), no significant spectral shift ($<$\,$1$\,km\,s$^{-1}$) was detected when cross-correlating the A and B spectra.

\begin{table*}[]
\caption[]{ Analysed spectroscopic observations of GJ\,436.}
\label{tab:observations}
\centering
\begin{tabular}{llllllc}
  \hline \hline
  \noalign{\smallskip}
  UT date & N$_{\rm exp}$ & T$_{\rm exp}$ & Airmass & Rel. hum & Mean S/N & Discarded orders \\
  &  & (s) &  & (\%) &  & \\
  \noalign{\smallskip}
  \hline
  \noalign{\bigskip}
  \textit{CARMENES} &  &  &  &  &  & \\
  \noalign{\smallskip}
  $2017$ Feb $02$ & $38$ & $280$ & $1.05$ & $69$ & $67$ & 63, 62, 57, 56, 54\,--\,51, 46, 45, 44, 43, 42, 40, 36 \\
  $2017$ Feb $17$ & $36$ & $272$ & $1.42$ & $70$ & $82$ & 63, 57, 55, 54, 53, 51, 50, 45, 44, 43 \\  
  $2018$ Jan $24$ & $84$ & $278$ & $1.22$ & $43$ & $78$ & 63, 57, 55, 54, 53, 51, 50, 45, 44, 43 \\
  $2018$ Feb $14$ & $82$ & $278$ & $1.22$ & $86$ & $61$ & 55, 54, 47, 46, 45, 44, 42 \\  
  $2018$ Apr $16$ & $31$ & $278$ & $1.10$ & $32$ & $90$ & 63, 57, 54, 53, 52, 45, 44, 43, 42, 41, 36 \\ 
  \noalign{\bigskip}
  \textit{CRIRES$^+$} &  &  &  &  &  & \\
  \noalign{\smallskip}
  $2023$ Jan $23$ & $90$ & $60$ & $1.66$ & $44.39$ & $204$ & None \\
  \noalign{\smallskip}
  \hline
\end{tabular}
\tablefoot{PI of the nights of 2017 Feb 02, 2017 Feb 17, and 2018 Apr 16 was L. Nortmann, while the PI for the night of 2023 Jan 23 was R. Landman. The nights of 2018 Jan 24 and 2018 Feb 14 were part of CARMENES Guaranteed Time Observations.}
\end{table*}

\section{Estimation of stellar and planetary parameters}
\label{sec:system_parameters}

\subsection{Stellar parameters}
\label{subsec:stellar_parameters}

We computed the stellar atmospheric parameters, namely $T_{\rm eff}$, $\log{g}$, and [Fe/H], on the CARMENES template spectrum corrected for telluric absorption with {\tt SteParSyn}\footnote{\url{https://github.com/hmtabernero/SteParSyn/}} \citep{Tabernero2022a} using the line list and model grid described by \cite{Marfil2021}. The luminosity was computed from the integration of the spectral energy distribution following \citet{Cifuentes2020} using  photometry from $B$ to $W4$ alongside the  \textit{Gaia} DR3 parallax \citep{GaiaDR3}. The stellar radius $R_\star$ was determined from the Stefan-Boltzmann law  using the {\sc SteParSyn} $T_{\rm eff}$ and the stellar mass $M_\star$ from the linear mass-radius relation given by \citet{Schweitzer2019}. Values for both stellar atmosphere parameters and derived stellar parameters are compiled in Table\,\ref{tab:stellar_params}. These values agree within $1\sigma$ with previous determinations \citep[][just to cite a few recent examples]{2021A&A...656A.162M,2022A&A...658A.194P,2022MNRAS.516.3802C,2024A&A...690A..58A,2024ApJ...973...90M}, but we preferred to use our own homogeneous determination of stellar and planetary parameters, as presented below.

\begin{table}[]
\renewcommand{\arraystretch}{1.3}
\centering
\caption{Calculated and selected parameters for GJ\,436.}
\label{tab:stellar_params}
\begin{tabular}{l c}
\hline
\hline
\noalign{\smallskip}
Parameter & GJ\,436 \\
\noalign{\smallskip}
\hline
\noalign{\smallskip}

\noalign{\smallskip}
\multicolumn{2}{c}{\textit{Stellar atmospheric parameters}} \\
\noalign{\smallskip}
$T_{\rm eff}\,$(K) & $3453\pm22$ \\
$\log{g}\,$(dex) & $4.74\pm0.07$ \\
$[\mathrm{Fe}/\mathrm{H}]\,$(dex) & $-0.19\pm0.10$ \\
\noalign{\smallskip}
\multicolumn{2}{c}{\textit{Derived stellar parameters} } \\
\noalign{\smallskip}
$R_\star\,(R_{\odot})$ & $0.43\pm0.02$ \\
$M_\star\,(M_{\odot})$ & $0.43\pm0.02$ \\
\noalign{\smallskip}
\hline
\end{tabular}
\end{table}

\subsection{Planetary parameters}
\label{subsec:planetary_parameters}

\subsubsection{TESS photometric data}
\label{subsubsec:tess_photometry}

The Transiting Exoplanet Survey Satellite (TESS) observed GJ 436 in 2 min short-cadence integrations in two different sectors, S22 and S49, in March 2020 and March 2022, respectively. All sectors were processed by the Science Processing Operations Center pipeline \citep[SPOC,][]{2016SPIE.9908E..1AC} and searched for transiting planet signatures with an adaptive wavelet-based transit detection algorithm  \citep{2010SPIE.7740E..0DJ}. In a first analysis of sector 22, GJ\,436 was announced on 15 April 2020 as a TESS object of interest (TOI) via the dedicated MIT TESS data alerts public website\footnote{\url{ https://tess.mit.edu/toi-releases/}} where the planet was identified at an orbital period of 2.64397$\pm$0.00003\,d. The planet was fitted with limb-darkening transit models by SPOC \citep{Li:DVmodelFit2019} and successfully passed all the diagnostic tests performed on them \citep{Twicken:DVdiagnostics2018}. No further transiting planet signatures were detected in the residual light curve by the SPOC. The planet candidate, GJ\,436\,b, has an estimated radius of 3.76$\pm$0.12\,R$_{\oplus}$ and a depth of 6480$\pm$56\,ppm.

We downloaded the light curve files for the two sectors from the Mikulski Archive for Space Telescopes. The first step was to verify that the simple aperture photometry (SAP) and  pre-search data conditioning SAP (PDCSAP) fluxes automatically computed by the pipeline are useful for scientific studies. For that, we checked that no additional bright source is contaminating the aperture photometry. Figure~\ref{fig:apertures} displays the target pixel files (TPFs) of GJ\,436 for the two sectors using the publicly available {\tt tpfplotter}\footnote{ \url{https://github.com/jlillo/tpfplotter}} code \citep{2020A&A...635A.128A}, which overplots the \textit{Gaia} Data Release 3 (DR3) catalogue on top of the TESS TPFs. We confirmed that there are no additional \textit{Gaia} sources within the photometric aperture around GJ\,436 automatically selected by the pipeline. Therefore, we considered the extracted TESS light curves to be free of contamination from nearby stars.

\subsubsection{HIRES, HARPS, and ESPRESSO spectroscopic data}
\label{subsubsec:hires_harps_spectroscopy}

GJ\,436\,b is a well-characterized planet that was initially detected with HIRES by \cite{2004ApJ...617..580B}. The planet has an orbital period of 2.64 days, a minimum mass of 23\,M$_\oplus$, and an eccentricity of 0.15. \cite{2007A&A...472L..13G} later confirmed it as a transiting planet with a size and mass comparable to Neptune. \cite{trifonov2018} combined 113 CARMENES precise Doppler measurements with those available in the literature from HIRES and HARPS to derived new orbital parameters for the system. The updated Keplerian parameters of GJ\,436\,b, based on the modelling of all 638 Doppler measurements, were a planet semi-amplitude $K_b$ = 17.38\,m\,s$^{-1}$, period of $P_b$ = 2.644 days, and eccentricity $e_b$ = 0.152. The inclination constraints from the transit ($i$ = 85.80$\pm$0.25\,deg) yield a planetary mass of $m_b$ = 21.4$\pm$0.2\,M$_\oplus$ and semi-major axis of $a_b$ = 0.028$\pm$0.001$\,$\,au. In our new analysis, we also included the ESPRESSO RV data used by \cite{2022A&A...663A.160B} to analyse the Rossiter–McLaughlin (RM) signal. We took advantage of the RM anomaly induced by GJ\,436\,b on the RV residuals measured during the two transits observed with ESPRESSO \citep[see Fig.~2 of][]{2022A&A...663A.160B} to exclude the RV phases affected by the RM throughout all the archival spectroscopic data collected here.

\subsubsection{CARMENES spectroscopic data}
\label{subsubsec:carmenes_spectroscopy}

We had additional CARMENES observations of GJ\,436 beyond those presented by \cite{trifonov2018}, aimed at refining the planetary mass measurement. We used a total of 176 high-resolution spectra from 2016 January 08 to 2024 May 25. Relative RVs were extracted separately for each \'echelle order using the {\tt serval} software \citep{2018A&A...609A..12Z}. The final VIS RVs per epoch were computed as the weighted RV mean over all \'echelle orders of the respective spectrograph, with an {\it rms} of 11.97\,m\,s$^{-1}$ and a mean error bar of 1.76\,m\,s$^{-1}$.

\subsubsection{Joint photometric and spectroscopic analysis}
\label{subsubsec:joint_analysis}

We derived the planetary parameters of GJ\,436\,b by performing a joint analysis of photometric and spectroscopic data, combining TESS light curves with RV measurements from HIRES, HARPS, CARMENES VIS, and ESPRESSO. The modelling was carried out using the {\tt juliet} Python package, which incorporates {\tt radvel} \citep{2018PASP..130d4504F} for Keplerian RV fitting and {\tt batman} \citep{2015PASP..127.1161K} for transit modelling. Outliers in the photometric data were filtered using a 3$\sigma$ clipping procedure.

Instead of directly fitting for the planet-to-star radius ratio ($R_{p}/R_{\star}$) and the impact parameter ($b$), we adopted the $r_{1}$ and $r_{2}$ parametrization introduced by \citet{2019MNRAS.490.2262E}, which allows both parameters to vary between 0 and 1 while fully covering the physically allowed space for $R_{p}/R_{\star}$ and $b$. We imposed a prior on the stellar density ($\rho_{\star})$, rather than on the scaled semi-major axis ($a$), ensuring a single consistent value of $\rho_{\star}$ across the system. To optimize computational efficiency, given the volume of TESS data points and the lack of additional transiting signals, we restricted the photometric dataset to $2.5$ hour windows centred on the predicted mid-transit times of GJ\,436\,b (with transits lasting approximately 1 hour). We defined normal priors for the parameters to the RV and light curve data $P$, $t_0$, $r_{1}$, $r_{2}$, centred on the values previously obtained from independent fits to the radial velocity and light curve data.

Given the generalized Lomb-Scargle periodogram of the RV data our joint model did not include a stellar activity component and omitted the use of a Gaussian process. Moreover, the expected RV semi-amplitude of the planet is sufficiently large that any potential stellar activity signal would have a negligible impact on the results. Based on the evidence reported in the literature, we modelled the system with one Keplerian signal: a 2.64-day transiting planet also detected in the RV data.

The posteriors from the one-planet fit that we adopted as planetary parameters for the GJ\,436 system are presented in Table \ref{tab:planets_params}. The folded light curves, with the two TESS sectors combined and phased with the orbital period of the transiting planet, are shown in Fig.~\ref{fig:tesslc}. The corner plot displaying the posterior distributions of some of the planetary parameters obtained from the joint fit is presented in Fig.~\ref{fig:gj436_cornerplot}. The resulting RV model folded in phase is displayed in Fig.~\ref{fig:gj436_RVmodel_vs_phase}.

The combined analysis of the TESS, HARPS, HIRES, ESPRESSO, and CARMENES data yields an orbital period of 2.6438972$\pm$0.0000004\,days. From the RV amplitude of 17.3$^{+0.16}_{-0.16}$\,$\rm m\,s^{-1}$ and an orbital inclination of $86.52^{+0.49}_{-0.45}$\,deg, we derived a true planetary mass of $21.02^{+0.69}_{-0.72}$\,M$_\oplus$, and radius of $3.96^{+0.19}_{-0.18}$\,R$_\oplus$ for the transiting planet GJ\,436\,b.

An additional analysis including the complete TESS light curves, fitting a GP simultaneously to detrend the light curves while modelling the transit signal, was performed. The obtained values were fully consistent with those reported in Table~\ref{tab:planets_params}, within 1$\sigma$ error bars. The planetary parameters previously reported in the literature \citep[e.g.][]{trifonov2018, maxted2022, 2022A&A...663A.160B} are also compatible within the 1$\sigma$ level.

\begin{table}[]
\renewcommand{\arraystretch}{1.3}
\centering
\caption{Calculated and selected parameters for GJ\,436\,b.}
\label{tab:planets_params}
\begin{tabular}{l c}
\hline
\hline
\noalign{\smallskip}
Parameter & GJ\,436\,b \\
\noalign{\smallskip}
\hline
\noalign{\smallskip}

\noalign{\smallskip}
\multicolumn{2}{c}{\textit{Fitted planet parameters}} \\
\noalign{\smallskip}
$P$ (d) & $2.6438972^{+0.00000038}_{-0.00000037}$  \\
$t_0$ & $2458902.31614^{+0.000073}_{-0.000070}$  \\
$K$ (m/s) & $17.3\pm0.16$  \\
$e$ & $0.160\pm0.0056$  \\
$\varpi$ & $327.9^{+2.7}_{-2.6}$  \\
$r_1$ & $0.903^{+0.0047}_{-0.0051}$  \\
$r_2$ & $0.0836^{+0.0017}_{-0.0013}$  \\

\noalign{\smallskip}
\multicolumn{2}{c}{\textit{Derived planet parameters} } \\
\noalign{\smallskip}
$R_{\rm p}/R_{\star}$ & $0.0836^{+0.0017}_{-0.0013}$ \\
$R_{\rm p}\,(R_{\oplus})$ & $3.96^{+0.19}_{-0.18}$  \\
$a/R_{\star}$ & $14.57\pm0.27$ \\
$a$ (au) & $0.0293\pm0.0014$ \\
$b = (a/R_{\star}) \cos i$ & $0.855^{+0.0070}_{-0.0076}$ \\
$i$ (deg) & $86.52^{+0.49}_{-0.45}$  \\
$t_{\rm 14}$ (h) & $0.998^{+0.035}_{-0.033}$ \\
$\delta_{\rm depth}$ (ppm) & $6991.0^{+290}_{-220}$ \\
$M_{\rm p} \sin i$ (M$_{\oplus}$) & $21.09^{+0.70}_{-0.72}$  \\
$M_{\rm p}$ (M$_{ \oplus}$) & $21.02^{+0.69}_{-0.72}$ \\
$\rho_{\rm p}$ (g $\rm cm^{-3}$) &  $1.87^{+0.30}_{-0.25}$ \\
$T_{\rm eq}$ (K) & $640.50^{+8.6}_{-8.4}$  \\
\noalign{\smallskip}
\hline
\end{tabular}
\end{table}

\section{Methods}
\label{sec:methods}

\subsection{Normalization, outliers, and masking}
\label{sec:methods_norm}

Ground-based observations of exo-atmospheres suffer from an inherent disadvantage that stems from the varying conditions of Earth's atmosphere (e.g. seeing and precipitable water vapour (PWV) variability). Such variations during the observations induce a fluctuating baseline in the spectra (Figs.\,\ref{fig:steps}a and \ref{fig:steps}b). Thus, to ensure a consistent level across all spectra, we performed for both CARMENES and CRIRES+ an order-by-order normalization by fitting a third-degree polynomial to the pseudo-continuum (see Fig.\,\ref{fig:steps}c).

Similar to the approach used by \cite{kesseli2021}, we removed contamination from cosmic rays by applying a $3\sigma$-clipping procedure where we flagged values that significantly deviate from the mean of the pixel's time series. These values were corrected by interpolating over their nearest neighbours in wavelength. Next, we masked the most opaque Earth's atmospheric transmittance (telluric) windows, where $80$\% of the normalized flux was absorbed for CARMENES. This value was relaxed to $85$\% in the case of CRIRES$^+$, following \citet{grasser2024}. A safety window of $5$ pixels was added to each flagged pixel to further limit potential contamination of strong telluric line wings. The percentage of masked pixels varies significantly for different spectral orders and it is shown, for each night, in Fig.\,\ref{fig:masked_pix}. For the CRIRES$^+$ dataset, we removed the first and last $100$ pixels of each spectral order.

\subsection{Telluric and stellar correction}

Telluric and stellar features within the spectra are orders of magnitude stronger than the Doppler-shifted excess absorption from the planetary atmosphere. In order to disentangle the exo-atmospheric signal, we used {\tt SysRem} \citep{tamuz2005, mazeh2007}, an iterative principal component analysis algorithm that accounts for unequal uncertainties for each data point. {\tt SysRem} has been successfully applied in similar studies in the past to reveal signatures in the atmosphere of several exoplanets \citep{de2013, birkby2013, birkby2017, sanchez2019, nugroho2020, cont2022, nortmann2025}. In addition, recent work by \cite{maguire2024} suggests that {\tt SysRem} yields a better performance for our purposes, compared to other common telluric-removal techniques \citep[e.g. {\tt Molecfit};][]{smette2015, kausch2015}.

Each spectral order is affected by different levels of contamination, essentially depending on the spectral shape of the telluric transmittance. As a result, the number of required {\tt SysRem} passes required is order-dependent and a criterion needs to be set to determine it. This is a difficult step that can lead to unintended biases and even to spurious planet-like signals, depending on the employed criterion \citep[see e.g.][]{cabot2019, cheverall2023}. Following the approach used by \cite{herman2020, herman2022, deibert2021, ridden2023, rafi2024, parker2025}, we studied the variance of the residual spectral matrices after applying {\tt SysRem} and halted the algorithm when the standard-deviation difference between two consecutive passes ($\Delta\sigma$) was below a given threshold ($1\%$) and started to plateau. This
metric therefore encapsulates the percentage change in the standard
deviation of the data for the application of each {\tt SysRem} iteration. Thus, the $\Delta\sigma$ metric is defined as

\begin{equation}
   \Delta \sigma=\frac{{ }^{(i-1)} \sigma-{ }^{(i)} \sigma}{^{(i-1)} \sigma},
\end{equation}

\noindent where $^{(i-1)} \sigma$ and $^{(i)} \sigma$ are the standard deviation of the residuals in a spectral order before and after the $i$-th pass of {\tt SysRem} respectively. In practice, this method helps recognize when $\Delta\sigma$ plateaus for each spectral order, indicating that the major spectral variations have been removed and no further passes are required. This method should preserve signals close to the noise level such as the exo-atmospheric absorption, while also being a model-independent approach, in contrast to injection recovery methods employed in the past \citep{alonso2019, sanchez2019, sanchez2022}.

\begin{figure}
    \centering
    \includegraphics[width=\columnwidth]{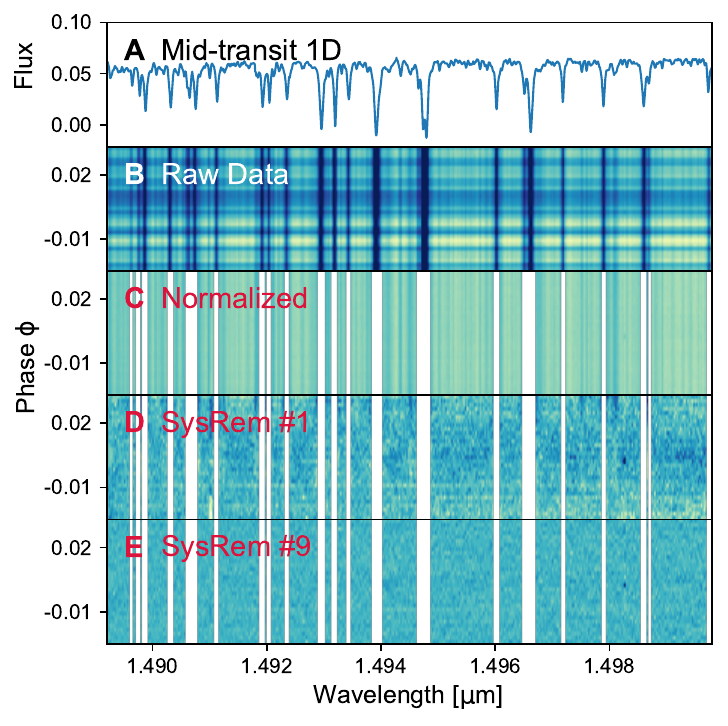}
    \caption{Steps of the data preparation applied to a representative spectral region of the dataset obtained on the night of February $04$, $2017$. The vertical axis in panel A represents flux in arbitrary units, while in all other panels it represents time, expressed as phase. Panel A: Original spectrum observed at mid-transit, where telluric absorption lines from H$_2$O can be identified. Panel B: Original spectral matrix extracted using \texttt{CARACAL}, where major S/N differences are observed between the spectra (horizontal stripes), and prominent telluric H$_2$O emission can be identified. Panel C: Normalized spectra, where a $3\sigma$ clipping has been applied to filter out outliers and where telluric masks have been applied (white, excluded pixels). All spectra are now normalized to a common continuum. Panel D: Resulting spectra after one {\tt SysRem} pass, where major residuals can be observed, especially from emission lines. Panel E: Spectra after nine {\tt SysRem} passes, where most of the telluric contribution has been removed and any exoplanet signal, if present, is buried in the noise.}
    \label{fig:steps}
\end{figure}

In the defined metric, a plateau is typically reached between two to five {\tt SysRem} passes for CARMENES datasets and five to seven for CRIRES$^+$ dataset, depending on the spectral order, as illustrated in Fig.\,\ref{fig:ds_per_order} and \ref{fig:ds_per_order_criresp}. During the first {\tt SysRem} passes, telluric and stellar features still remain (panel D of Fig.\,\ref{fig:steps}), suggesting the need for further cleaning. 
For intermediate passes, the exo-atmospheric signal,if present, is expected to be buried in the noise (Fig.\,\ref{fig:steps}E).

\subsection{Model atmospheres}
\label{subsec:model_atmospheres}

We modelled the exo-atmospheric transmission spectra using \texttt{petitRADTRANS}\footnote{\textcolor{magenta}{\texttt{https://petitradtrans.readthedocs.io}}} ({\tt pRT}), a Python-based package for calculating transmission and emission spectra of exoplanets through radiative transfer \citep{molliere2019}. We assumed H$_2$-He dominated atmospheres where Rayleigh scattering and collision-induced absorption are included \citep{figueira2009, nettelmann2010}. The same preparation steps performed on the real data were reproduced for the models to ensure that the same distortions are introduced.

\begin{figure}
    \centering
    \includegraphics[width=\columnwidth]{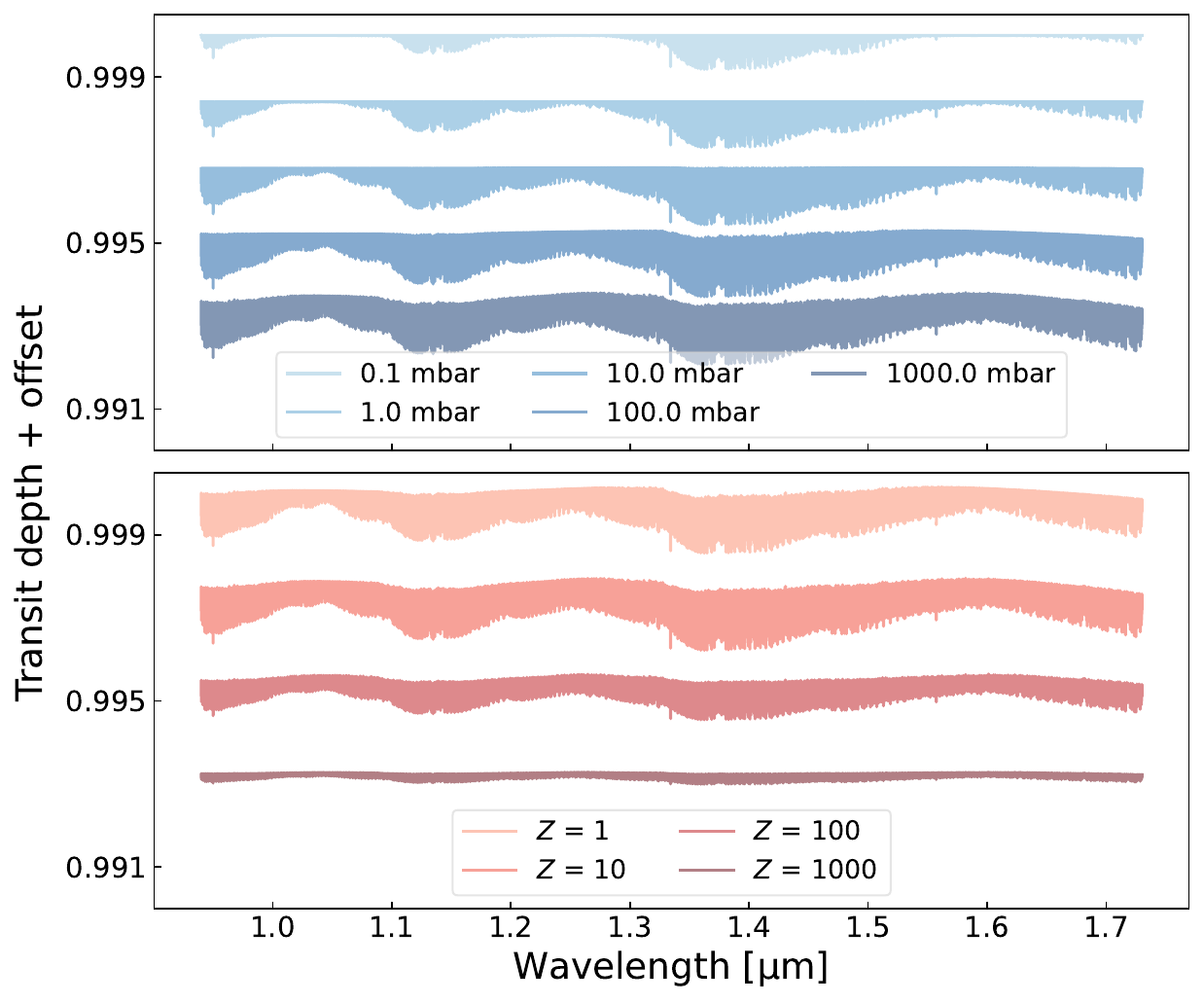}
    \caption{Transit depth of H$_2$O for GJ\,436\,b generated with \texttt{pRT} as a function of cloud deck pressure level (top panel, fixed metallicity of $10\times$\,solar) and atmospheric metallicity (bottom panel, clear atmosphere) as multiples of solar. For clarity, an offset of $0.002$ in transit depth was applied to the transmission models in both panels. Clouds at higher altitudes mute more H$_2$O lines.}
    \label{fig:clouds}
\end{figure}

We used {\tt pRT} to model the absorption of the most relevant spectroscopically active species in the covered NIR range. Specifically, we considered H$_2$O and CO using the line lists presented by \cite{rothman2010}, NH$_3$ by \cite{sousa2014}, and CH$_4$ by \cite{hargreaves2020}. Model transmission spectra were generated on a two-dimensional grid spanning the pressure level of a potential cloud deck ($p_{\rm c}$) and the metallicity of the atmosphere ($Z$). This latter step was facilitated by the algorithm \texttt{easyCHEM}\footnote{\textcolor{magenta}{\texttt{https://easychem.readthedocs.io}}} \citep{lei2024}, which provides with the mass fractions of the atmospheric compounds and the mean molecular weight (thus, the atmospheric scale height) as a function of the metallicity. Our $p_{\rm c}$ grid spanned from $0.1$\,mbar to $1$\,bar, while the $Z$ grid covered from $1\times$\,solar to $1000\times$\,solar metallicity, assuming a solar C/O ratio of $0.55$ \citep{asplund2009}. Grid values were evenly spaced on a logarithmic scale, resulting in a total of $100$ templates per molecule. Whereas only the respective species' opacity is used to compute the transmission spectra in {\tt petitRADTRANS}, \texttt{easyCHEM} does consider multiple species to assess the atmospheric compositions and scale height.
Figure\,\ref{fig:clouds} illustrates the variations of the transit depth in the models due to changes in both $p_{\rm c}$ and $Z$.

\subsection{Cross-correlation technique}
\label{subsec:cross-correlation}

\begin{figure}
    \centering
    \includegraphics[width=\columnwidth]{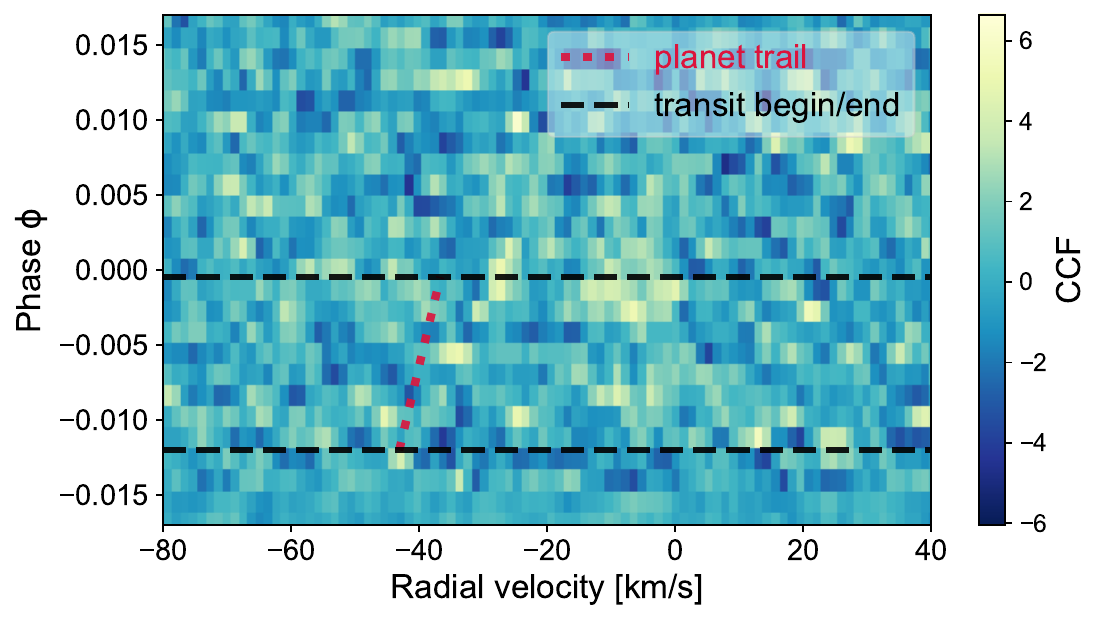}
    \caption{Cross-correlation analysis of potential H$_2$O signals in the transmission spectrum of GJ\,436\,b observed with CARMENES in the near-infrared on the night of February $02$, $2017$. 
    We show the cross-correlation matrix, CCF ($v$,$i$), in the Earth's rest frame as a function of the velocity Doppler shifts applied to the template (horizontal axis) and the planet's orbital phase (vertical axis). The cross-correlation was derived using a $10\times$\,solar metallicity template with a cloud deck at $10$\,mbar, co-adding all useful spectral orders. Dashed-dotted black horizontal lines mark the start and end of the transit, while the dotted red line indicates the expected exoplanet velocities with respect to Earth, accounting for eccentricity. 
    \label{fig:cc_water}}
\end{figure}

Even after removing the telluric and stellar contributions, any potential atmospheric lines from GJ\,436\,b remain buried below the noise level. By using the cross-correlation technique, it is possible to co-add the contribution from potentially hundreds of these lines in the form of a cross-correlation function (CCF) peak \citep[e.g.][]{snellen2010, brogi2012, de2013, birkby2013, birkby2017}. We performed the cross-correlation between the residual matrices obtained in the previous steps $R_{ij}$ and the transmission models (templates) $m_j$ calculated using {\tt pRT}. In order to explore a wide range of velocities with respect to the Earth, we Doppler-shifted the templates in a range of velocities $v$ ($-325$ to $+325$\,km\,s$^{-1}$) with respect to the Earth's rest frame by linear interpolation. We used $1.3$\,km\,s$^{-1}$ (CARMENES) and $1$\,km\,s$^{-1}$ (CRIRES$^+$) intervals, as set by the average velocity step size between the instruments pixels. CCFs were obtained individually for each spectrum, $i$, forming a cross-correlation matrix, $CCF(v,i)$, as follows:

\begin{equation}
\label{eq:ccf_vi}
    CCF(v,i) = \sum_j \sum_\lambda \frac{R_{i,\,j}(\lambda)\,m_j(\lambda,\,v)}{\hat \sigma_{i,\,j}^2 (\lambda)},
\end{equation}

\noindent where $\hat \sigma (\lambda)$ are the propagated uncertainties of the data and the summations loop over wavelength ($\lambda$) and spectral order $j$, so as to co-add the potential information contained in our full spectral coverage. The resulting total $CCF(v,i)$ in the Earth's rest frame is illustrated in Fig.\,\ref{fig:cc_water}, where the trace of GJ\,436\,b would be expected to appear during transit (i.e. between the horizontal dash-dotted red lines) along the planetary RVs with respect to the Earth, $v_{\rm p}$. To properly describe the particular orbit of GJ\,436\,b, we modified the classic definition of $v_{\rm p}$ to account for an eccentric orbit as follows:

\begin{equation}
    v_{\mathrm{p}}(\phi)=K_{\mathrm{p}} \cdot [\cos(\nu(\phi) +  \omega_{\rm p})+e \cdot \cos(\omega_{\rm p})]-v_{\text {bary }}+v_{\text {sys}},
    \label{eq:vp_eccentric}
\end{equation}

\noindent where $\phi$ is the orbital phase, $\nu(\phi)$ is the true anomaly, $\omega_{\rm p}$ the argument of periastron, $e$ is the eccentricity, $v_{\rm bary}$ the barycentric velocity due to Earth's motion around the Solar System's barycentre, $v_{\rm sys}$ the systemic velocity of the star-planet system, and $K_{\rm p}$ the radial velocity semi-amplitude of the planet, defined as

\begin{equation}
    K_{\mathrm{p}}=\frac{2 \pi a}{P_{\rm orb} \sqrt{1-e^2}} \sin(i),\end{equation}

\noindent where $a$ is the orbital parameter (the semi-major axis), $P_{\rm orb}$ the orbital period, and $i$ the inclination.

The eccentric $v_{\rm p}$ has an average shift of about $20$\,km\,s$^{-1}$ and a slight change in slope compared to the circular velocity. This change highlights the importance of accurate orbital characterization, as failing to account for the correct orbital configuration could result in incorrect interpretations of potential signals \citep{basilicata2024, grasser2024}. It is at this point that we visually inspected the cross correlation matrices in the Earth's rest-frame for all spectral orders and for each night separately. We discarded from the analysis those orders where the telluric contamination was not properly removed by {\tt SysRem} (last column of Table\,\ref{tab:observations}). That is, when a trail of high CCF values along the vertical at $0$\,km/s is observed (i.e. where correlation with uncorrected telluric-H$_2$O lines is expected to appear). We used the same list of H$_2$O-based discarded orders for all CCF analysis, regardless of the studied molecule, as a conservative approach to minimize telluric noise and opaque wavelength windows as much as possible. As it is the usual approach, we also excluded spectral orders where the target molecule does not have significant spectral features \citep[e.g.][]{parker2025}.

Next, we co-added all in-transit CCFs along the time axis (i.e. sum over $i$ in Eq.\,\ref{eq:ccf_vi}) to maximize any potential signal. To do this, we first Doppler-shift $CCF(v,i)$ to the exoplanet's rest-frame by using Eq.\,\ref{eq:vp_eccentric}, assuming $K_{\mathrm{p}}$ as unknown and using a range of values from $-300$ to $300$\,km\,s$^{-1}$. Thereby, an absorption with a planetary origin should only appear around the expected $K_{\mathrm{p}}$ of GJ\,436\,b (i.e. the true exoplanet rest frame) in the form of a CCF peak.

\begin{figure}
    \centering
    \includegraphics[width=\columnwidth]{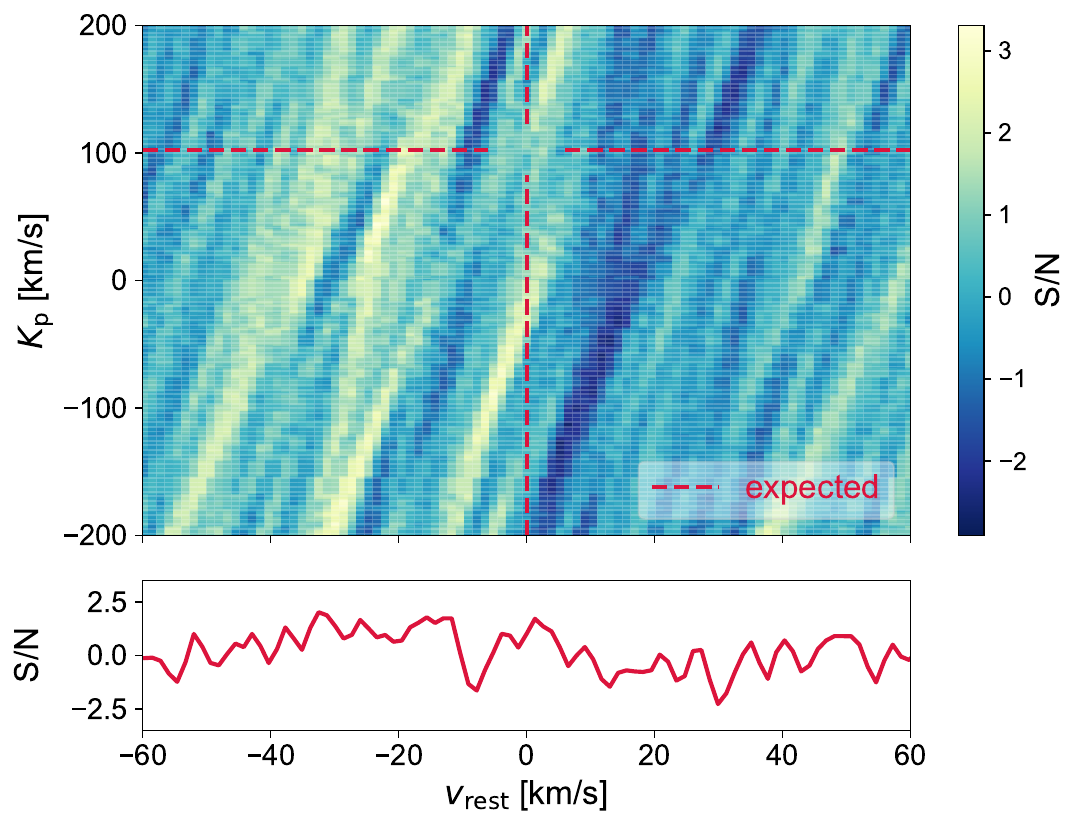}
    \caption{Significance analysis for potential water vapour cross-correlation signals on the night of February $02$, $2017$. Top panel: Signal-to-noise ratio map as a function of radial velocity in the exoplanet's rest frame ($v_{\rm rest}$, horizontal axis) and the projected orbital velocity semi-amplitude, $K_{\rm p}$ (vertical axis). Horizontal red lines denote the expected $K_{\rm p}$, and vertical red lines mark the expected zero velocity, assuming eccentric orbits. Bottom panel: One-dimensional cross-correlation function at the expected $K_{\rm p}$ ($102.3$\,km\,s$^{-1}$) of GJ\,436\,b for the eccentric case.
    \label{fig:cc_water2}}
\end{figure}

\subsection{Assessment of potential signals}
\label{subsec:significances}

\subsubsection{S/N for CCF studies}

Following common signal significance evaluation approaches, we computed a S/N map as a function of the exoplanet rest-frame velocity and $K_{\rm p}$ (top panel of Fig.\,\ref{fig:cc_water2}; \citealt{birkby2017, brogi2018, alonso2019, sanchez2019, nugroho2021, cont2022, cont2024}). For each $K_{\rm p}$, we divided each CCF point by the CCF's standard deviation, excluding a $\pm 20$\,km/s window around it. The division ensured that we avoided inflating the standard deviation by including potential CCF-peak wings. Potential signals are hence evaluated in a common $K_{\rm p}$--$\rm v_{rest}$ map, as shown in the top panel of Fig.\,\ref{fig:cc_water2}. In this case, the map does not reveal any significant signals (e.g. at S/N$>$\,$4$) at the expected $K_{\rm p}$ (bottom panel of Fig.\,\ref{fig:cc_water2}). All the observed patterns could be well explained by noise fluctuations or weak telluric residuals. This then leads to a non-detection of water vapour in GJ\,436\,b when using this particular CARMENES dataset observed on February 02, 2017. The same methods are applied subsequently to all transit datasets (Fig.\,\ref{fig:snr_individual}).

To enhance the recovery of a potential planetary signal, we combined the information from each observed night by creating a merged CCF map. In this process, we sorted the frames from all datasets along the time axis according to their corresponding orbital phases following the approach of \cite{cont2022, cont2025}. The respective $v_{\rm bary}$ of each spectrum was considered at the time of Doppler-shifting the CCF matrix from the Earth's rest-frame to the planet's. This procedure hence yielded a single CCF map, which incorporates all the data from the different nights and is used to compute the $K_{\rm p}$--$v_{\rm rest}$ and S/N maps in the same way as for a single-night dataset.

\subsubsection{Bayesian retrieval framework}

To investigate the multidimensional parameter space of the transmission models, we employed the nested sampling algorithm {\tt MultiNest} \citep{feroz2013}, along with its Python wrapper {\tt PyMultiNest} \citep{buchner2016}, to sample a three-dimensional space defined by the logarithm of the atmospheric metallicity in units of the solar value, $\log(Z/Z_{\odot})$, the logarithm of the pressure level of the cloud deck, $\log(p_{\rm c})$, and the uncertainties-scaling parameter $\beta$. 
We adopted uniform priors on all free parameters, as summarized in Table\,\ref{tab:priors}, with a broad coverage of plausible atmospheric conditions. 
We set per-night retrievals with $300$ live points each and a non-constant efficiency mode \citep[e.g.][]{blain2024}. These parameters and prior ranges served as coordinates for the algorithm, which assesses the goodness-of-fit of the model to the data by evaluating a log-likelihood function $\log(\mathcal{L})$. Here, we followed the framework of \citet{gibson2020, gibson2022} and \citet{maguire2024}, with
\begin{equation}
    \log \mathcal{L}(F \mid \theta) = - \frac{1}{2} \sum_{N,\,\lambda,\,t} \left(\frac{\mathcal{P}(F(\lambda,\,t)) - \mathcal{P}(M_\theta\,(\lambda,\,t))}{\beta\,\hat \sigma(\lambda,\,t)} \right)^2
    - N \log(\beta),
\label{eq:loglike}
\end{equation}
where $\theta=\left[ \log(Z/Z_\odot),\log_{10}(p_c),\beta\right]$ are the parameter vectors, $\mathcal{P}$ represents a mathematical operator including all operations performed to the data in order to enable exo-atmospheric detections (i.e. preparations steps such as normalization and telluric correction), $F$ represents our data spectral matrix, $M_\theta$ represents the model template computed from the parameters $\theta$, $\beta$ is a scaling parameter for the uncertainties, and where the sum loops over all useful nights $N$, spectral points and orders ($\lambda$), and in-transit spectra ($t$). The same preparation steps performed on the real data were applied to the model so as to ensure the same distortions were introduced in both sets, hence increasing their likeness. Consequently, this method determines the best-fitting model to the data using Bayesian statistics \citep{brogi2019, gibson2020, gibson2022, cont2022, blain2024}.

Importantly, when running the retrieval, we slightly changed the normalization step described by Sect.\,\ref{sec:methods_norm} to follow exactly the same steps described in \citet{gibson2022}, which we summarize below. This change was to ensure that no statistical bias is introduced, while performing a significantly faster normalization step in all $\log(\mathcal{L})$ evaluations. In practice, this entails starting from the outlier-corrected CARMENES data, performing a normalization dividing each order by the median spectrum over time, and applying {\tt SysRem} order by order following the criterion discussed above. For the determined number of {\tt SysRem} passes for each order, we stored the resulting {\tt SysRem} column vectors $U$ and precomputed the {\tt SysRem} -filtering projector $\mathcal{P}$ that was used to prepare models before comparing them to the data \citep{gibson2020, gibson2022, cont2025}. Essentially, a filtered model is obtained by dividing it by its median ($\mathcal{M_{\rm norm}}$), and then computing
\begin{equation}
    \mathcal{M_{\rm filt}} = \mathcal{P}(\mathcal{M_{\rm norm}}) = U(\Lambda U)^{\dagger}(\Lambda \mathcal{M_{\rm norm}}),
\label{eq:m_filt}
\end{equation}

\noindent where $\Lambda$ is the inverse of the mean $\sigma$ over wavelength, and where the matrix of {\tt SysRem} column vectors has an additional column of ones to account for potential offsets introduced by media division at the time of normalizing the data. Since the projector $\mathcal{P} = U(\Lambda U)^{\dagger}\Lambda$ is precomputed, each retrieval iteration is greatly accelerated. Essentially, each likelihood evaluation entails a call to {\tt easyCHEM} to provide the atmospheric mass fractions and mean molecular weight, a call to {\tt pRT} to obtain the transmission spectrum with a given cloud contribution and including the opacities from H$_2$, He, H$_2$O, CH$_4$, NH$_3$, CO, and CO$_2$, followed by a preparation of the templates as per Eq.\,\ref{eq:m_filt}, and the subsequent log-likelihood computation from Eq.\,\ref{eq:loglike}.

Each night's dataset was retrieved independently to produce weighted posterior samples and per-night multivariate kernel density estimates (KDEs; non-parametric estimators of the posterior constructed from the weighted samples). For each night we computed the marginalized one- and two-dimensional posteriors as usual, where the posterior is formally defined as
\begin{equation}
    \log P(\theta \mid F) = \log \mathcal{L}(F \mid \theta) + \log \pi(\theta) - \log \mathcal{Z},
\end{equation}
with $\mathcal{L}(F\mid\theta)$ the likelihood of the data $F$ given parameters $\theta$, prior $\pi(\theta)$, and $\mathcal{Z}$ the Bayesian evidence. To obtain the combined posteriors for all nights, we built a candidate set of parameter vectors $\theta$ by drawing from the per-night KDEs, from the priors, and from the pooled per-night samples. Subsequently, we used the KDEs to evaluate the per-night posterior densities for each $\theta$, denoted $p_i(\theta)$, and combined them in log-space according to\begin{equation}
    \log w(\theta) = \sum_{i=1}^{N_{\rm nights}} \log p_i(\theta) - (N_{\rm nights}-1)\,\log \pi(\theta),
\end{equation}
which corrects for the prior being included in each per-night posterior. The weights $w(\theta)$ were normalized and used to resample the candidate pool, producing combined samples from which medians and $68\%$ credible intervals ($16$th, $50$th, and $84$th percentiles) and the final corner plot were derived. To ensure broad coverage of parameter space, the candidate pool combined KDE draws, prior draws, and pooled per-night samples. This procedure approximates the joint posterior for the combined dataset while preserving per-night posterior structure and avoiding prior double-counting.

This combination implicitly assumes that different nights are statistically independent (so the joint likelihood factorizes), have the same parameter definition, prior ranges, and functional form for the model and preparation pipeline (even if individual models or {\tt SysRem} projectors differ for different spectral orders and nights), and that the per-night KDEs adequately approximate the night-wise posteriors. Due to this, it was possible for us to combine the retrieval information for $\log(Z/Z_\odot)$ and $\log_{10}(p_c)$, but this approach was not used for the $\beta$ parameter, since it explicitly reflects per-night noise properties. Thus, we opted for presenting the retrieval of $\beta$ separately for each night in Sect.\,\ref{sec:retrievals}.

\begin{table}
  \centering
  \caption{Prior ranges and posterior medians ($68$\% credible intervals).}
  \label{tab:priors}
  \begin{tabular}{lccc}
    \hline
    \hline
    \noalign{\smallskip}
    Free parameters                      & Prior range & Retrieved        \\
    \hline
    \noalign{\smallskip}
    $\log_{10} (Z/Z_{\odot})$            & $\mathcal{U}(-2,3)$           & $0.3$\,$\pm$\,$1.3$           \\
    $\log_{10}(p_{\rm c})$ (bar)     & $\mathcal{U}(-9,3)$    & $-5.1$\,$\pm$\,$2.1$                \\
    $\beta_{CARM,\,1}$                         & $\forall \beta,\,\mathcal{U}(10^{-2},\,10^2)$ & $0.7764$\,$\pm$\,$0.0006$             \\
    $\beta_{\rm CARM,\,2}$                         &    &$0.7820$\,$\pm$\,$0.0005$            \\
    $\beta_{\rm CARM,\,3}$                         &   &$0.7981$\,$\pm$\,$0.0006$            \\
    $\beta_{\rm CARM,\,4}$                         &   &$0.7995$\,$\pm$\,$0.0005$             \\
    $\beta_{\rm CARM,\,5}$                         &   &$0.7580$\,$\pm$\,$0.0005$            \\
    $\beta_{\rm CRIRES}$                         & & $1.4320 \pm 0.0005$            \\
    \noalign{\smallskip}
    \hline
    \noalign{\smallskip}
    Fixed parameters         & Value     & Unit            \\
    \hline
    \noalign{\smallskip}
    $K_P$             & $102.3$   & km\,s$^{-1}$    \\
    $v_{\rm wind}$    & $0.0$     & km\,s$^{-1}$    \\
    \hline
    \noalign{\smallskip}
  \end{tabular}
  \tablefoot{Posterior medians and uncertainties are reported as median $\pm$ the symmetric half-width of the central 68\% credible interval, i.e.\ $1\sigma \equiv 0.5\,(P_{84}-P_{16})$, where $P_{16}$, $P_{50}$, and $P_{84}$ are the 16th, 50th (median), and 84th percentiles of the posterior. Values reported for $\log_{10} (Z/Z_{\odot})$ and $\log_{10}(p_{\rm c})$ were obtained from posterior combination of all nights. The five $\beta_{\rm CARM,i}$ and $\beta_{\rm CRIRES}$ entries correspond to the per-night parameters.}
\end{table}

\section{Results and discussion}
\label{sec:results_discussion}

We studied the presence of H$_2$O, CH$_4$, CO, and NH$_3$ in the atmosphere of GJ\,436\,b by applying the cross-correlation technique to five useful CARMENES datasets and one from CRIRES$^+$. In the following, we present the results of direct CCF, CCF-based injection recovery of artificial signals and Bayesian retrievals.

\subsection{Non-detection of molecular signals with CCF}
\label{subsec:non-detection}

After performing the CCF analysis with an atmospheric model with $p_{\rm c}$\,=\,$10$\,mbar and $Z$\,=\,$10\times$\,solar, we did not detect any signal of H$_2$O, CH$_4$, or CO in GJ\,436\,b's atmosphere with either CARMENES or CRIRES$^+$ (Fig.\,\ref{fig:all_mol_sum}).
We additionally searched for the expected weak spectral features of NH$_3$, but no signals were obtained either (not shown).
From this result and previous knowledge of the existence of a H$_2$\,--\,He atmosphere in this planet, we can think of three possible scenarios: (i) GJ\,436\,b does not have significant amounts of H$_2$O, CH$_4$, or CO in its atmosphere (which seems unlikely), (ii) our assumed model is significantly different from the real conditions in the atmosphere, a problem that is tackled by using Bayesian retrievals in Sect.\,\ref{sec:retrievals}, or (iii) the atmosphere is veiled by high-altitude clouds that suppress spectral features to the extent that even high-resolution spectroscopy cannot discern the line cores. We assess this point by discussing the results from different methodologies.

\subsection{Upper limits from injection recovery tests}
\label{subsec:inj_rec_tests}

\begin{figure*}
    \centering
    \includegraphics[width=\textwidth]{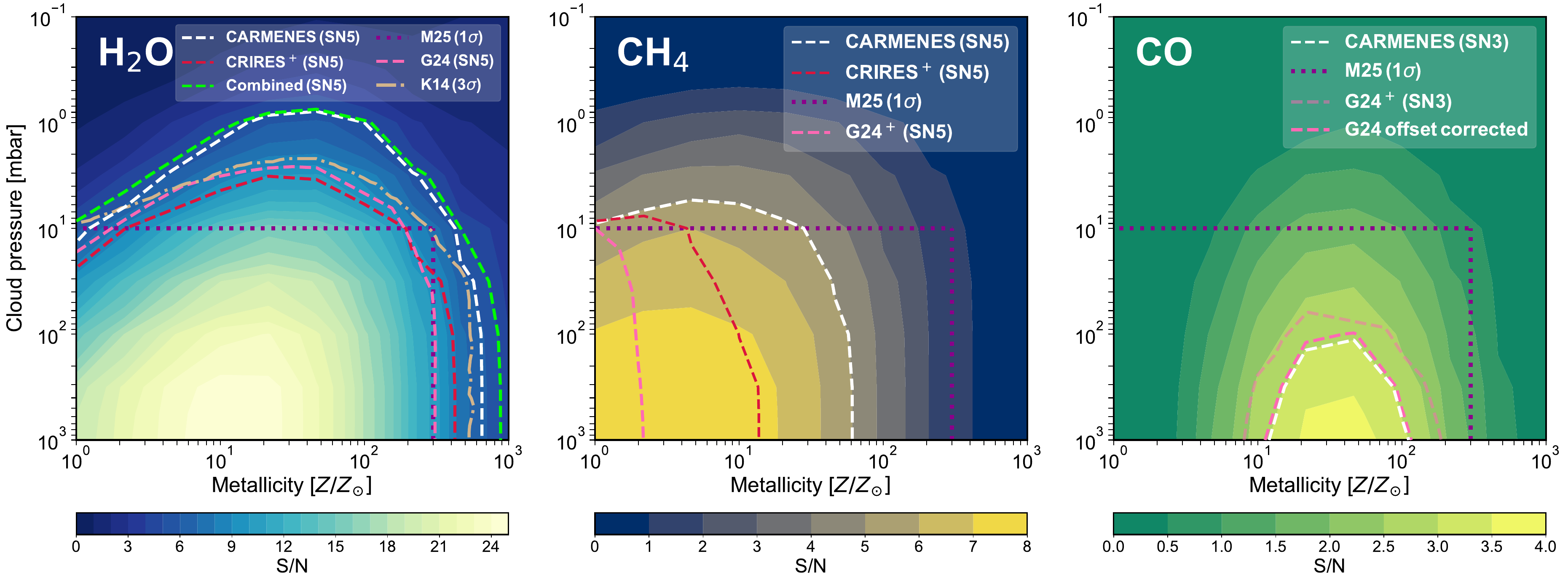}
    \caption{Detectability maps obtained from injection recovery tests, showing upper limits on the abundances of the main near-infrared absorbers in the atmosphere of G\,436\,b. Specifically, we investigated H$_2$O (left panel), CH$_4$ (middle panel), and CO (right panel) across a range of metallicities and cloud deck altitudes. Dashed lines indicate the S/N\,$=$\,$5$ contours (`SN5', S/N\,$=$\,$3$ for CO denoted as `SN3') obtained from CARMENES (white, five transits combined), CRIRES$^+$ (red, one transit), and combining the results from both instruments (green).
    The dotted pink lines mark the S/N\,$=$\,$5$ (S/N\,$=$\,$3$ for CO) contours from \cite{grasser2024}, referred to as `G24', using the same CRIRES$^+$ dataset. For CO, an additional contour with a potential offset corrected is also shown as `G24 offset corrected' (see discussion in Sect.\,\ref{subsec:carmenes_perf}). The dash-dotted brown line corresponds to the $99.7$\% upper limit from Fig.\,3 in \cite{knutson2014a}, referred to as `K14'. The latter values were extracted from their respective figures. The dashed purple line represents the $68$\% upper limit placed by \cite{mukherjee2025}, referred to as `M25'. The area below each curve represents the discarded atmospheric parameter values.}
    \label{fig:injections}
\end{figure*}
    
In the spectral range studied, H$_2$O and CH$_4$ exhibit prominent absorption features in the form of line forests, whereas CO shows only a relatively weak band around \SI{1.5}{\micro\meter} with far fewer lines. As we did not detect the presence of any of these molecules in our data we proceeded to place limits on their detectability. To this end, we performed signal-injection tests using a grid of H$_2$O, CH$_4$, and CO templates with varying $p{\rm_c}$ from $0.1$\,mbar to $1$\,bar, and $Z$ ranging from $1\times$\,solar to $1000\times$\,solar. The model signals were injected before the normalization step at the expected velocities of the exoplanet with respect to the Earth (Eq.\,\ref{eq:vp_eccentric}). We then applied our preparation and CCF methods for each injected case, and used the same model as a template for cross-correlation in that grid point. The resulting S/N of the recovered signals are shown in Fig.\,\ref{fig:injections}. We note that, in this process, the non-injected CCF is subtracted from the injected one to remove any potential noise structures at the injected position, hence ensuring the significance represented better the recovery of the injected atmosphere.

Focusing on the H$_2$O detectability map (Fig.\,\ref{fig:injections}, left), which is the most restrictive case of the three molecules, we combined the five CARMENES transits and determined that the most plausible scenario for GJ\,436\,b's atmosphere is the presence of clouds at pressures lower (higher altitudes) than the $10$\,mbar level if the metallicity is low ($1$--$2\times$\,solar), below the $1$\,mbar level for intermediate metallicities ($10$--$100\times$\,solar), or cloud-free above a metallicity of $\sim$\,$600\times$\,solar. For any other case, our detection thresholds indicate that, if real, we should have observed hints of H$_2$O absorption in our analysis.
Interestingly, we observed CARMENES-derived upper limits (after combining transits) to be significantly tighter than the particular single-transit results from CRIRES$^+$. We note that our CRIRES$^+$ upper limits for water vapour agree rather well with the previous analysis of this dataset by \cite{grasser2024}. The difference between the results of both instruments for H$_2$O and the other species is further explored in Sect.\,\ref{subsec:carmenes_perf}.

When studying CH$_4$ (Fig.\,\ref{fig:injections}, middle), we find CARMENES constraints to be significantly stronger than those derived with CRIRES$^+$, and both CH$_4$ curves are generally less stringent than our H$_2$O upper limits. This does not come as a surprise, since methane’s intrinsically weaker lines result in a larger fraction of the parameter space being undetectable. 
However, we find significantly more restrictive constraints than reported in \citet{grasser2024} when studying the same CRIRES$^+$ dataset, a discrepancy that is addressed in Sect.\,\ref{subsec:carmenes_perf}.
From CARMENES results, CH$_4$ upper limits discard cloudless or low-altitude cloud ($p_c$\,$>$\,$6$\,mbar) scenarios for any metallicity $<60\times$\,solar.

Regarding CO (Fig.\,\ref{fig:injections}, right), we performed our analysis only for CARMENES and in the useful spectral orders where this molecule shows absorption lines (CARMENES orders $39$ and $38$), which are also covered by CRIRES$^+$ orders $4$ and $5$. Since the CO injection-recovery tests were already performed for CRIRES$^+$ by \citet{grasser2024}, and due to these H-band CO lines being relatively weak, we opted for not repeating this study using CRIRES$^+$. Indeed, from Fig.\,\ref{fig:injections}, a detection threshold of S/N\,=\,$5$ is not met for any combination of $p_{\rm c}$\,--\,$Z$, and we could only show the S/N\,=\,$3$ contours as a common reference.
At first glance, we observed an apparent difference in sensitivity, with CRIRES$^+$ constraints being slightly more stringent than CARMENES'. However, when slicing Fig.\,\ref{fig:injections} (right panel) at different $Z$ and $p_{\rm c}$ values \citep[i.e. reproducing Fig.\,$6$ from][]{grasser2024}, we observed a consistent offset of about $0.4$ between the baseline S/N levels of our CO signals and theirs (Fig.\,\ref{fig:slices_comparison}). We were able to reproduce a nearly identical offset with the CARMENES data when using all spectral orders overlapping with the full CRIRES$^+$ H1567 coverage. That is, when including also the spectral orders with no CO lines (which were originally discarded in our analysis). The inclusion of all CRIRES$^+$ H1567 orders in \citet{grasser2024} has been confirmed with the authors (Grasser; priv. comm.). After considering this offset (see ``G24 offset corrected curve''), we found that the CO detection limits are nearly identical in both studies (see Sect.\,\ref{subsec:carmenes_perf}), but only allowed us to exclude a region with metallicities between $10\times$\,solar and $100\times$\,solar, with cloud decks ranging from $1000$ to $100$\,mbar. In other words, CO studies did not improve the H$_2$O constraints either.

Overall, our CCF injection recovery tests are dominated by the most-restrictive water vapour case. We find that our results are consistent with the high-altitude cloud deck and/or high metallicity scenarios inferred by \cite{figueira2009}, \cite{moses2013}, \cite{knutson2014a}, \cite{grasser2024}, and \cite{mukherjee2025} for GJ\,436\,b's atmosphere.

\subsection{CARMENES and CRIRES$^+$ combined}
\label{subsec:carmenes_criresp_comb}

The tightest possible constraints on the detectability of GJ\,436\,b’s atmosphere using cross-correlation are obtained by combining CRIRES$^+$ and CARMENES observations. For this, we restricted our analysis to water vapour, as it is the case that yields the most stringent upper limits (Fig.\,\ref{fig:injections}, green).

Interestingly, the combined constraints with both instruments are very similar to those placed with CARMENES only. This might indicate that the signal-detectability in GJ\,436\,b's atmosphere has a physical limit imposed by $p_{\rm c}$. That is, if the planet's atmosphere has clouds at altitudes higher than the $1$\,mbar-level (which is a sensible assumption given ours and previous results), a few additional transit datasets with current instrumentation might not be help us disentangle molecular lines that could be almost completely muted. Similarly, the detectability of H$_2$O becomes extremely challenging or unfeasible both at the solar and $1000\times$ solar metallicity boundaries, due to either its low abundance or highly compressed atmospheres. Hence, we would not expect these upper limits to significantly change until the arrival of the next-generation of ground-based telescopes and high-resolution spectrographs.

\subsection{A study of instrumental capabilities}
\label{subsec:carmenes_perf}

The different constraints between our CARMENES and CRIRES$^+$ analyses are a combination of multiple factors, involving different telescope apertures ($3.5$\,m for CAHA/CARMENES versus $8.2$\,m for the VLT/CRIRES$^+$), instrumental properties (e.g. resolving power or total wavelength coverage), and observing conditions (e.g. different challenges involving the telluric correction).
The first of these can be compensated, in principle, by combining several CARMENES transits. Despite the challenges of preparing multiple datasets, this allowed us to build a similar S/N to that of a single CRIRES$^+$ observation at the VLT. That is, a direct comparison between single-transit datasets from CARMENES and CRIRES$^+$ would provide significantly worse results for the former, given that the VLT does collect over five times more photons per exposure.

After combining all CARMENES observations, the broader spectral coverage of its NIR channel, compared to a single CRIRES$^+$ setting, does indeed contribute substantially to obtaining more restrictive results. 
In order to illustrate this, we repeated the injection recovery tests shown in Fig.\,\ref{fig:injections} for H$_2$O and CH$_4$, but including only the CARMENES orders that approximately overlap with those of the CRIRES$^+$ H1567 setting (see Figs.\,\ref{fig:range_comparison} and \ref{fig:range_comparison_ch4_co}), obtaining the injection-recovery maps shown in Fig.\,\ref{fig:injections_7orders}. This approach, while still hampered by differences in other instrumental properties and observing conditions, allowed us to evaluate the impact of H$_2$O and CH$_4$ absorption bands in the $0.96$\,--\,$1.45$\,$\mu$m range.
For H$_2$O, the results from Fig.\,\ref{fig:injections_7orders} show that CARMENES-derived upper limits worsened and became weaker than those of CRIRES$^+$ at the low- and high-metallicity cases, as expected from discarding many useful spectral points. The more restrictive upper limit from CRIRES$^+$ in these cases is likely due to a combination of its higher resolution ($\mathcal{R}$\,=\,$100\,000$) being able to resolve weaker H$_2$O cores at low $Z$, and more compressed atmospheres at high $Z$. However, we observe nearly identical constraints for intermediate metallicities (from $10$ to $100\times$\,solar) and $\sim$\,$3$\,mbar cloud decks, for which the water vapour content is significant and both instruments disentangle the signal with a similar performance (albeit combining multiple CARMENES observations).
In the case of CH$_4$, we found a similar effect: CARMENES upper limits became more relaxed when reducing the spectral points included in the analysis, bringing the new S/N\,$=$\,$5$ curve closer to CRIRES$^+$ results. In this case, CARMENES-restricted constraints were still slightly more stringent. Since both instruments cover the strongest methane lines in the H-band between $1.6$ and $1.7\,\mu$m with almost overlapping wavelength coverage, the origin of the difference is challenging to determine. Given the similar (though slightly worse) results obtained for H$_2$O, a potential explanation might be a more difficult telluric removal in five different datasets for CARMENES, with respect to only one for CRIRES$^+$ 
This issue might then be relaxed when studying CH$_4$, revealing a slightly better performance for CARMENES. However, providing definite evidence for this hypothesis is beyond the scope of this work.

Interestingly, we found the order-restricted CARMENES results to be now more similar to our CRIRES$^+$-derived upper limits for CH$_4$, but both remained notably more restrictive than the curve presented in \cite{grasser2024}. Since we used the same CRIRES$^+$ dataset and given the very consistent results for H$_2$O \citep[between our analyses and the upper limits in][]{grasser2024}, the origin of the difference is somewhat uncertain. Lastly, for CO, the CARMENES spectral coverage does not introduce additional spectral lines than those covered by CRIRES$^+$ (see Fig.\,\ref{fig:range_comparison_ch4_co}) and thus, the results of Fig.\,\ref{fig:injections} already depict comparable results for both instruments \citep[after considering the potential CO offset in][]{grasser2024}.

\subsection{Bayesian retrievals}
\label{sec:retrievals}

\begin{figure*}
\centering
\includegraphics[scale=0.55]{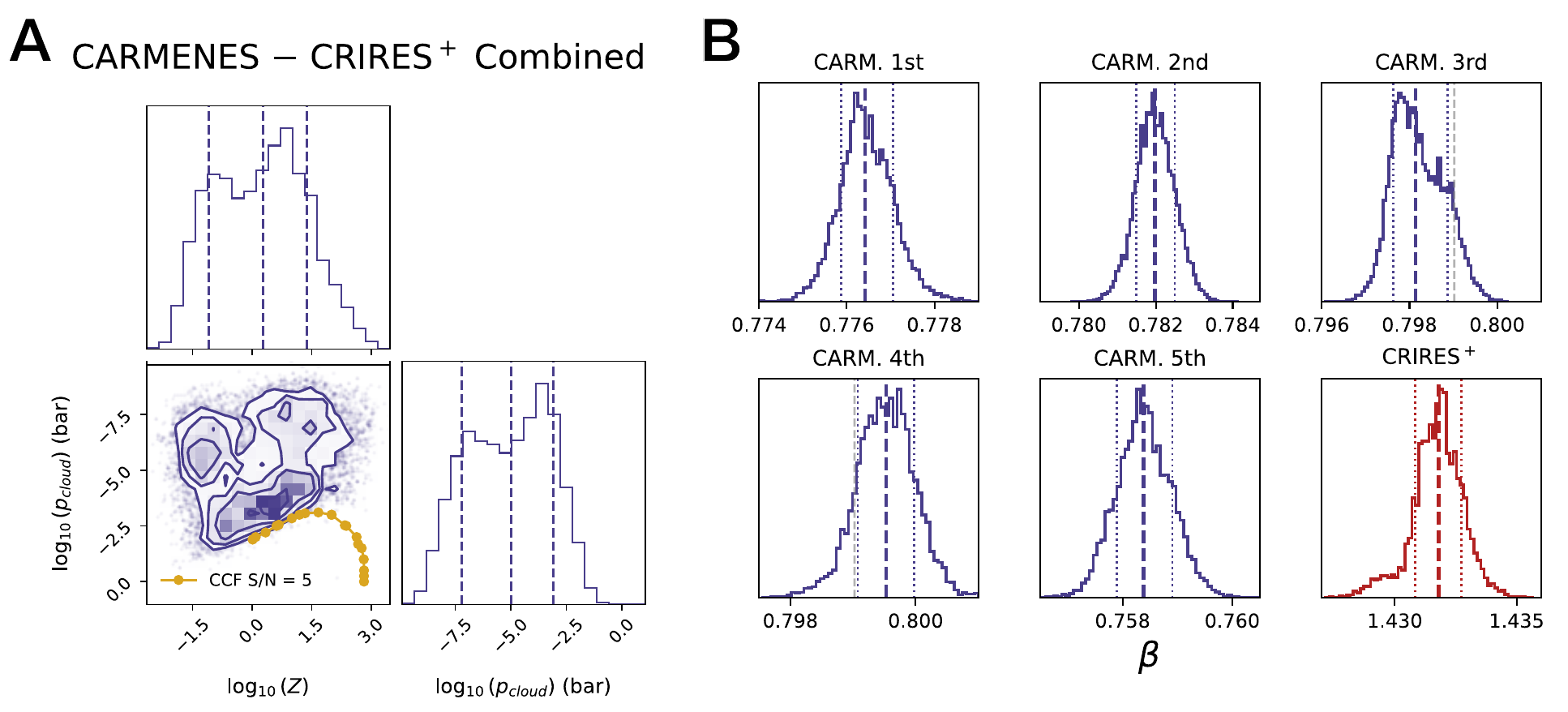}
\caption{Posterior distributions from atmospheric retrievals of six transit datasets of GJ\,436\,b, five from CARMENES, and one from CRIRES$^+$.
Panel A: Combination of all useful transit datasets from both instruments. This yields the marginalized one-dimensional posteriors shown in the diagonal panels for $\log_{10}(Z/Z_{\odot})$ and $\log_{10}(p_{\rm c})$, with vertical lines at the $16$th, $50$th, and $84$th percentiles. The off-diagonal panel shows the two-dimensional posterior density estimated from the combined samples, with contours indicating enclosed-probability regions (we display 68.3\%, 95.5\%, and 99.7\% levels). For comparison, the CCF injection-recovery curve (S/N\,$=$\,$5$, yellow) is overplotted in the $\log_{10}(p_{\rm c})$ vs.\ $\log_{10}(Z/Z_{\odot})$ panel. Panel B: Marginalized one-dimensional posteriors for the $\beta$ parameter separately for CARMENES (five transits) and CRIRES$^+$ (one transit).}

\label{fig:retrievals_full}
\end{figure*}

We performed a Bayesian retrieval of our six transit datasets obtained with CARMENES and CRIRES$^+$, exploring a wider parameter space than with direct CCF and using a different metric to verify our previous results (see also Table\,\ref{tab:priors}). With Fig.\,\ref{fig:retrievals_full}, we confirm the non-detection of species in the atmosphere of GJ 436\,b with Bayesian retrievals, as best-fitting models present primarily a low-pressure (high-altitude) cloud deck (pressures\,$<$\,$10$\,mbar) that mute any potential feature regardless of the atmospheric metallicity. For high-metallicity scenarios, at $\log_{10}(Z/Z_{\odot})$\,$\gtrsim$\,$500\times$ solar, the atmospheric scale height becomes too small, and the transit signal is undetectable in our data regardless of the presence of clouds. For low and intermediate metallicities, we observe a general correlation between $\log_{10}(Z/Z_{\odot})$ and $\log_{10}(p_{c})$: the higher the former is, the higher the altitude of the cloud deck needs to be to mute stronger H$_2$O lines. In other words, this means our retrieval only finds good fits to the combined datasets when no prominent spectral lines are present in the templates, reflecting that no exo-atmospheric lines are detected in the data.

The retrieval thus ratified the constrains put by CCF-based injection recovery tests. In Fig.\,\ref{fig:retrievals_full} we overplot the CCF injection–recovery curve corresponding to a detection threshold of S/N\,$=$\,$5$. The region lying beneath that curve thus defines combinations of metallicity and cloud-top pressure for which our CCF pipeline would have produced a detectable signal, while the posterior region returned by the Bayesian retrieval encloses those parameter combinations that are consistent with the observed spectra (i.e. effectively no-signal scenarios). These two metrics are therefore complementary and mutually reinforcing: the retrieval-allowed area occupies parameter space that lies predominantly outside the CCF-detectable domain, meaning that the models consistent with the data would not have been picked up by our CCF search at S/N\,$=$\,$5$. 

Regarding the uncertainties-scaling parameter, we find it to range between $0.75$ and $0.80$ for CARMENES data (see Table\,\ref{tab:priors}), which is in line with previous retrievals of this instrument using this or similar frameworks \citep{yan2022, guo2024, blain2024, cont2022, cont2025}. This means that we find only a slight overestimation of uncertainties by {\tt CARACAL}, or slight overcorrection of spectral features by our {\tt SysRem} -based preparation pipeline. However, a retrieval on the CRIRES$^+$ dataset reveals a $\beta_{\rm CRIRES}= 1.432 \pm 0.001$, which indicates a larger discrepancy between the pipeline uncertainties we derived and the scatter in the prepared CRIRES$^+$ data. This suggests either an underestimation of noise by the standard error propagation in our pipeline or additional residual systematics not being captured by the model (e.g. telluric residuals). Indeed, when preparing the data using the methods of \citet{gibson2022} prior to the retrieval, we observed some telluric residuals. However, the retrieved $\beta_{\rm CRIRES}$ did not decrease significantly when increasing the number of {\tt SysRem} passes. Hence, it is likely that the issue might be a combination of both some telluric residuals and slight underestimation of the uncertainties by our raw-data reduction pipeline.

\subsection{Detectability of H$_2$O with ANDES}
\label{sec:ANDES}

The absence of molecular signatures in our CARMENES and CRIRES$^+$ analyses motivated an assessment of what next-generation facilities might reveal for a planet like GJ\,436\,b. To this end, we generated a suite of simulated ELT/ANDES observations using {\tt EXoPLORE}, an end-to-end high-resolution spectroscopic simulator, based on the methods described by \cite{blain2024}\footnote{A separated publication and an open-access framework for {\tt EXoPLORE} are in development.}. This framework makes use of {\tt easyCHEM} and {\tt petitRADTRANS} for the chemical modelling of the atmosphere and radiative transfer computations, respectively. The tool produces synthetic, high-resolution time-series spectra that incorporate the planned $YJH$-bands wavelength coverage and S/N estimates derived from the ANDES exposure time calculator \citep[ETC, integrations of 30\,s, see][]{sanna2024andes}, a resolving power of $\mathcal{R}$\,$=$\,$100,000$, and a parametrised treatment of telluric evolution with airmass. The current implementation does not yet include additional sources of time-correlated noise nor the impact of rapid PWV fluctuations. Hence, the detectability values reported here should be interpreted as optimistic upper limits on the achievable performance.

For consistency with our observational analysis, we explored a two-dimensional grid in atmospheric metallicity and cloud-top pressure. For each grid point ($\log{Z/Z_\odot}$,\,$p_{\rm c}$), {\tt EXoPLORE} generated a complete synthetic transit dataset that includes fundamental spectral contributions, namely:
\begin{itemize}
    \item We evaluated the telluric absorption using transmission curves from ESO’s \texttt{Skycalc} Tool \citep{noll2012atmospheric, jones2013advanced}, at the airmass of each synthetic exposure.
    \item The modelled exoplanet contribution containing H$_2$, He, H$_2$O, CO$_2$, CO, CH$_4$, NH$_3$, and HCN was injected after applying the appropriate orbital Doppler shift (Eq.~\ref{eq:vp_eccentric}) and was scaled by the instantaneous transit depth computed with {\tt BATMAN} \citep{kreidberg2015batman}, adopting a uniform limb-darkening profile and the orbital geometry of GJ\,436\,b (Table\,\ref{tab:planets_params}).
    \item The stellar spectrum was taken from the PHOENIX library \citep{husser2013new}, selecting a template consistent with the properties of GJ\,436, T$_\mathrm{eff}$\,=\,$3600$\,K, $\log{g}$\,=\,$4.5$, [Fe/H]\,$=$\,$+0.5$, and  [$\alpha$/M]\,$=$\,$+0.2$, closely following (within the database choices) the values reported in \citet{mukherjee2025} and our own values (Table\,\ref{tab:stellar_params}).
    \item Spectral noise was randomly generated for each wavelength and frame, using Python's functions {\tt np.random.default\_rng(seed\,$=$\,12345)} and {\tt rng.normal()}, where the standard deviation was computed for each bin as $\sigma(\lambda)$\,$=$\,$1/\text{(S/N)}_{\rm ETC}(\lambda)$. 
\end{itemize}

\begin{figure}
\centering
\includegraphics[width=\columnwidth]{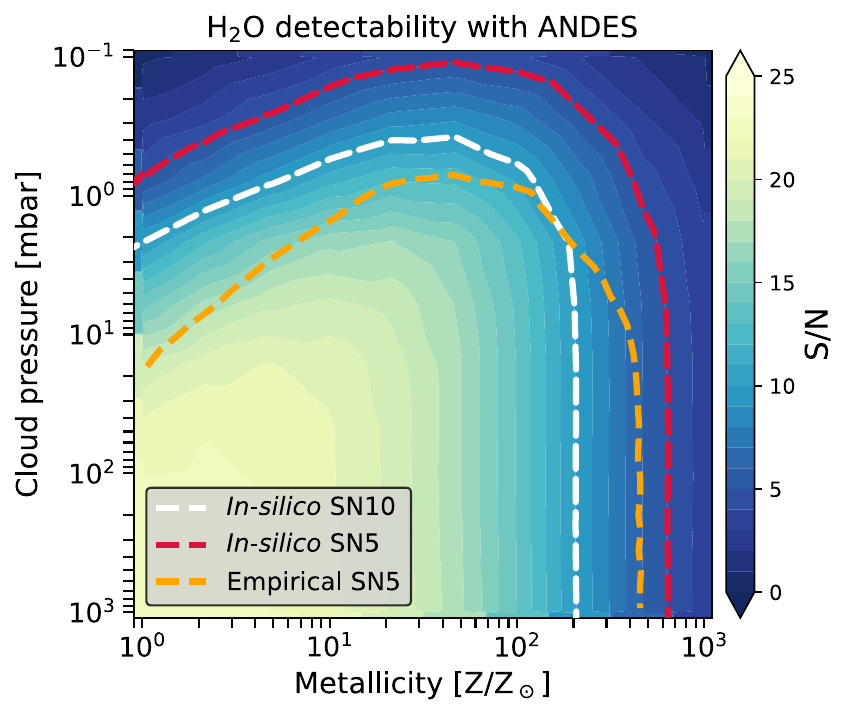}
\caption{Simulated H$_2$O detectability map for GJ\,436\,b produced with {\tt EXoPLORE} assuming a single transit observed with ANDES. The white and red contours indicate recovered S/N levels of 10 and 5 with \textit{in-silico} data, respectively. To illustrate the improvement over current constraints, we show the  S/N\,$=$\,$5$ contour (orange curve) derived from combining the CARMENES and CRIRES$^+$ data (Fig.\,\ref{fig:injections}). Although based on simulations, the map illustrates the significant improvement in sensitivity that ANDES and the ELT are expected to provide for probing cloudy sub-Neptune atmospheres.}
\label{fig:ANDES}
\end{figure}

The resulting \textit{in-silico} time series of spectra were then processed through the same cross-correlation pipeline used for the real observations (Sect.\,\ref{subsec:cross-correlation}) and the recovered signal at the expected $K_P$\,--\,$v_{\rm wind}$ was recorded as the predicted detectability for that atmospheric scenario.
Figure\,\ref{fig:ANDES} presents the resulting detectability map for water vapour assuming a single transit of GJ\,436\,b, also comparing the \textit{in-silico} predictions with the empirical results presented in this work. 

The simulations show that the large collecting area of the ELT would provide a substantial boost in sensitivity relative to current facilities. The predicted significance is highest, as expected, in the region already discarded by the data analysed here, where faint cloud opacity (cloud tops at a few millibars or deeper) yield H$_2$O detections at S/N\,$\gtrsim$\,10 from a single transit. The major improvement we observed came hence from the sensitivity to weak signals originating above clouds, as we would expect to detect H$_2$O in atmospheres for $p_c$\,$>$\,$0.1$\,mbar and $10$ \,$<$\,$Z$\,$<$\,$100\times$ solar at a S/N\,$\sim$\,$5$, which is almost an order-of-magnitude improvement over current instrumentation. However, the map reflects that atmospheres with extreme metal enrichment will remain challenging, since the reduced scale height suppresses the H$_2$O signal.

Therefore, our simulations indicate that ELT/ANDES will substantially enhance the recovery of molecular features in low mean-molecular-weight atmospheres, where the transmission signal is still in the photon-limited regime. In such cases, the increased collecting area boosts the visibility of line cores emerging above high-altitude clouds.
For atmospheres with very high metallicities though, the compressed scale
height still leads to weak line core absorption (see, also, Fig.\,\ref{fig:clouds}). Thus, even the ELT
would require multiple transits to secure a detection. This is also consistent with both the ELT simulation (one transit) and the analysed data combination (five CARMENES transits plus one CRIRES$^+$ dataset) yielding similar significances at intermediate and high-metallicities (i.e. at the signal-limited regime). Nevertheless, the ability to probe high-metallicity atmospheres by combining multiple ELT datasets represents a new avenue for probing these exoplanets, one that is not reachable with current facilities.

\section{Conclusions}
\label{sec:conclusion}

We present the analysis of six transit datasets of the warm Neptune GJ\,436\,b  observed with the high-resolution spectrographs CARMENES (five events) and CRIRES$^+$ (one event) at NIR wavelengths. After applying the CCF technique, performing injection recovery tests, and running Bayesian retrievals, we were able to establish the most stringent constraints on the detectability of H$_2$O, CH$_4$, and CO in the atmosphere of GJ\,436\,b from ground-based observations to date. The main conclusions derived from our work are:

\begin{enumerate}
    
    \item Our CCF studies of the CARMENES and CRIRES$^+$ datasets did not recover any significant signals for the studied species in this exoplanet's atmosphere, which is in line with findings by \citet{knutson2014a}, \citet{grasser2024}, and \citet{mukherjee2025}. Nevertheless, molecular species such as H$_2$O, CH$_4$, NH$_3$, and CO are expected in GJ\,436\,b's atmosphere, which is thought to be metal rich \citep{moses2013, knutson2014a}.

    \item We performed CCF-based injection recovery studies so as to better assess the detectability of atmospheric compounds with our combination of instruments, nights, and techniques. We ruled out atmospheric scenarios for GJ\,436\,b with low-altitude ($p_{\rm c}$<\,$1$\,mbar) or no cloud deck coverage and medium to low metallicities ($Z$<\,$600\times$\,solar), as such atmospheres would have been detectable in our datasets. These values agree well with the upper limits derived by models \citep{moses2013} and observations at high and low spectral resolution \citep{knutson2014a, grasser2024, mukherjee2025}. The most likely explanation for the lack of observable spectral features in the atmosphere of GJ\,436\,b is the presence of a cloud deck at higher altitudes than $<$\,$1$\,mbar pressure level, and/or a very high atmospheric metallicity, beyond $600\times$\,solar.

    \item A joint analysis of five CARMENES transits of GJ\,436\,b provides tighter constraints for H$_2$O and CH$_4$ than a single CRIRES$^+$ observation, essentially due to the broader coverage of the former instrument. For CO, however, both datasets yield similar (high) constraints since they cover the same isolated CO band. Restricting the CARMENES orders to match CRIRES$^+$ H1567 coverage resulted in CARMENES upper limits being similar, albeit slightly less restrictive, to those obtained with CRIRES$^+$. This occurs especially at the lowest and highest metallicities explored (i.e. weaker H$_2$O lines), potentially due to the higher resolving power.

    \item The tightest CCF-based upper limits to date in the atmosphere of GJ\,436\,b are obtained by combining all available CARMENES and CRIRES$^+$ datasets and performing injection recovery tests of H$_2$O, the strongest absorber in the NIR range covered by both CARMENES and CRIRES$^+$. This analysis determines that the atmosphere presents a cloud deck at altitudes higher than the $1$\,mbar level (regardless of the atmospheric metallicity), higher than the $10$\,mbar level for a solar metallicity, or presenting extremely high metallicities ($\gtrsim 900\times$ solar), regardless of cloud presence.

    \item Bayesian retrievals of the six transit datasets confirm that no prominent atmospheric features are detectable in GJ\,436\,b. The combined posteriors show that only low-pressure (high-altitude) cloud decks or very high-metallicity atmospheres are consistent with the data. The regions allowed by the retrieval fall outside the CCF-detectable domain, which means that the data effectively rule out models that would produce detectable spectral signals. Therefore, these two metrics provide a mutually consistent picture of an essentially featureless transmission spectrum.

    \item Our ELT/ANDES simulations indicate that the leap to 40\,m-class telescopes will significantly expand the atmospheric parameter space accessible for sub-Neptune characterization, even in the presence of clouds. For GJ\,436\,b, molecular features arising above cloud decks at pressures of $0.1$\,--\,$1$\,mbar could be detectable in a single transit across a broad metallicity range ($10$\,$<$\,$Z$\,$<$\,$300\times$ solar), which is unattainable with present-day instruments. At the same time, atmospheres with high metal enrichment ($Z$\,$>$\,$400\times$ solar) may remain intrinsically difficult targets due to their compressed scale heights, and meaningful detections in that regime are likely to require the combination of multiple ELT transits. Taken together, these results highlight both the new opportunities and the persistent physical limits that will shape atmospheric studies in the ELT era.

    \item For completeness, we derived precise and homogeneous stellar and planetary parameters from TESS, CARMENES, HIRES, HARPS, and ESPRESSO data, which are consistent with with previous determinations but add extra robustness to our work.
    
\end{enumerate}

These findings confirm GJ\,436\,b among the class of sub-Neptunes with intrinsically challenging transmission spectra. Looking ahead, our ELT/ANDES simulations demonstrate that the forthcoming generation of $40$\,m-class facilities will open up a substantial new discovery space and potentially enable routine access to molecular features that remain beyond the reach of current instruments. As such, GJ\,436\,b will become a valuable benchmark for testing the capabilities of next-generation observatories and for deepening our understanding of the atmospheric diversity of warm Neptunes.

\begin{acknowledgements}

We thank the anonymous referee for their very helpful comments on this manuscript. We also thank Natalie Grasser for kindly sharing her results with us for comparison purposes.
CARMENES is an instrument at the Centro Astron\'omico Hispano en Andaluc\'ia (CAHA) at Calar Alto (Almer\'{\i}a, Spain), operated jointly by the Junta de Andaluc\'ia and the Instituto de Astrof\'isica de Andaluc\'ia (CSIC).

 Funding for CARMENES has been provided by the
 Max-Planck-Institut f\"ur Astronomie (MPIA),
 Consejo Superior de Investigaciones Cient\'{\i}ficas (CSIC),
 European Regional Development Fund (ERDF),
 Ministerio de Ciencia, Innovaci\'on y Universidades (MICIU),
 Deutsche Forschungsgemeinschaft (DFG)
 and the members of the CARMENES Consortium
 (\url{https://carmenes.caha.es}).

 We acknowledge financial support from the Agencia Estatal de Investigaci\'on (AEI/10.13039/501100011033) of the MICIU and the ERDF ``A way of making Europe'' through projects
 PID2022-141216NB-I00,
 PID2022-137241NB-C4[1:4],
 PID2021-125627OB-C3[1:2],
 RYC2022-037854-I,
and the Centre of Excellence ``Severo Ochoa'' and ``Mar\'ia de Maeztu'' awards to the Instituto de Astrof\'isica de Andaluc\'ia (CEX2021-001131-S) and Institut de Ci\`encies de l'Espai (CEX2020-001058-M).

 This work was also funded by the
Centres de Recerca de Catalunya (CERCA),
a Fraunhofer-Schwarzschild fellowship at the Ludwig-Maximilians-Universit\"at M\"unchen and DFG under Germany's Excellence Strategy (EXC 2094--390783311),
and the European Union through European Research Council (ERC) Consolidator grant EvapourATOR (No. 101170037).
Views and opinions expressed are those of the author only and do not necessarily reflect those of the European Union or the ERC. Neither the European Union nor the granting authority can be held responsible for them.

\end{acknowledgements}

\bibliographystyle{aa}
\bibliography{biblio}

\onecolumn

\begin{appendix}

\section{Additional plots}
\label{app:additional_plots}

\begin{figure*}[h!]
    \centering
    \includegraphics[width=\textwidth]{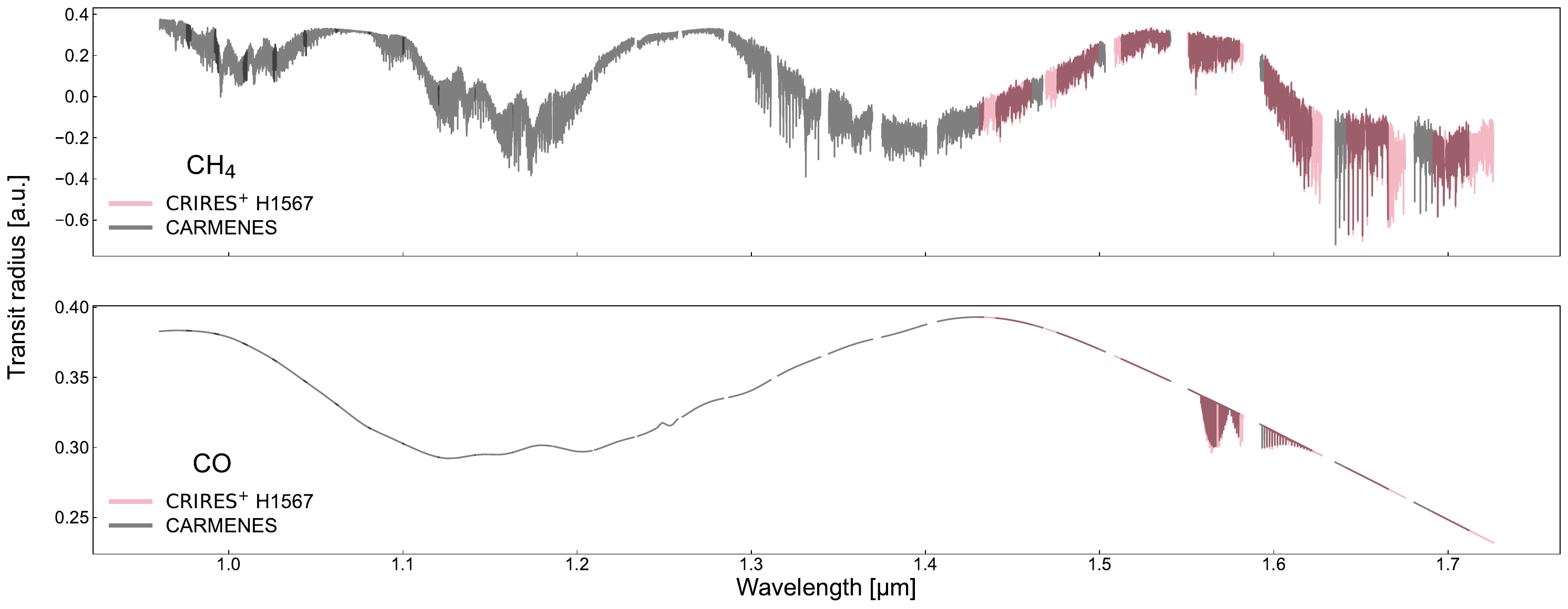}
    \caption{Spectral coverage comparison of CH$_4$ (top) and CO (bottom) transmission spectrum models for GJ\,436\,b, as observed by the CARMENES NIR channel (grey) and the $\mathrm{CRIRES}^{+}$ H1567 setting used by \citet[red]{grasser2024}.}
    \label{fig:range_comparison_ch4_co}
\end{figure*}
    
\begin{figure}[h!]
        \centering
        \includegraphics[width=0.35\textwidth]{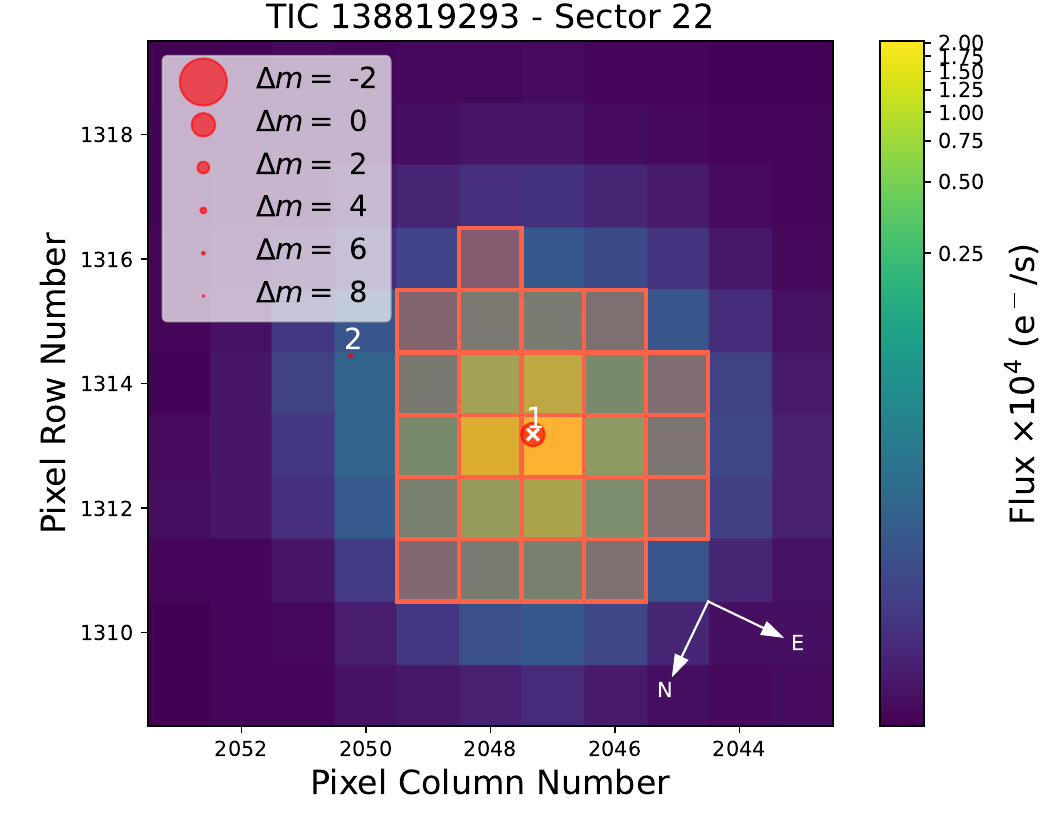}
        \includegraphics[width=0.35\textwidth]{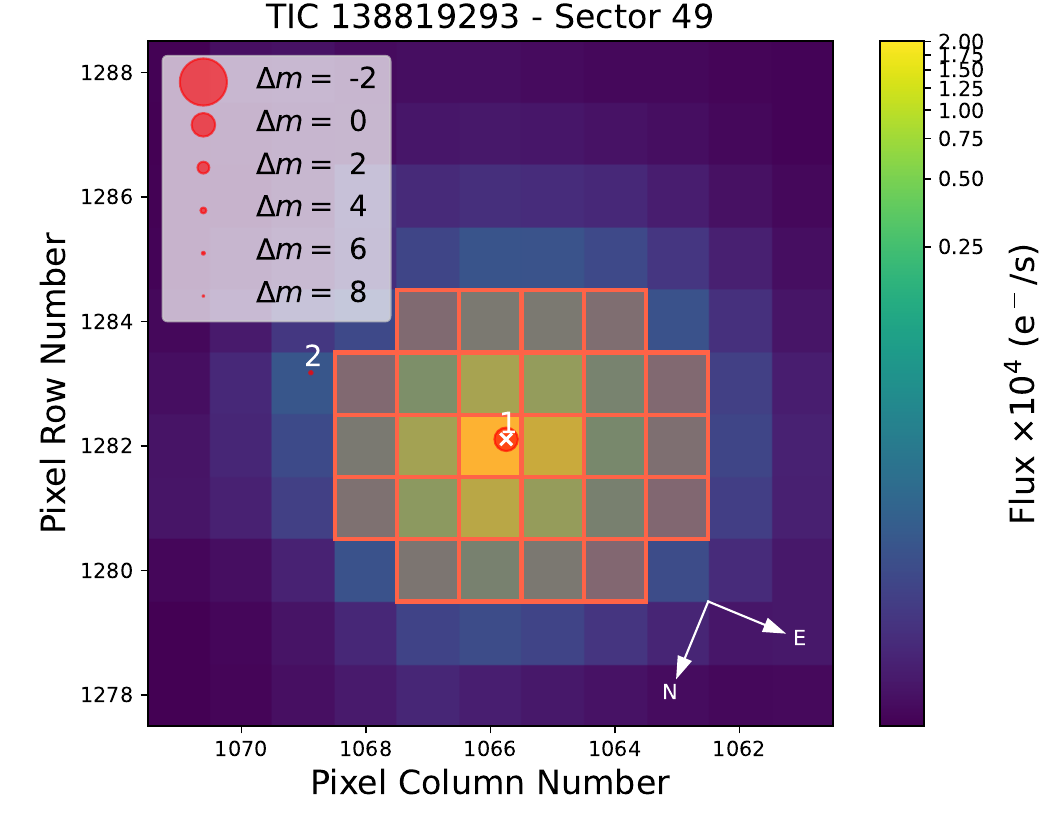}
        \caption{TPF of GJ\,436 (cross symbol) in TESS Sectors 22 (left) and 49 (right). The electron counts are colour-coded. The TESS optimal photometric aperture per sector used to obtain the SAP fluxes is marked with red squares. The \textit{Gaia} DR3 sources with $G$-band magnitudes down to 8\,mag fainter than GJ\,436 are labeled with numbers (GJ\,436 corresponds to number 1) and their scaled brightness based on \textit{Gaia} magnitudes is shown by red circles of different sizes (see figure inset). The pixel scale is 21\,arcsec\,pixel$^{-1}$.}
        \label{fig:apertures}
\end{figure}

\begin{figure}[h!]
        \centering
        \includegraphics[width=0.45\columnwidth]{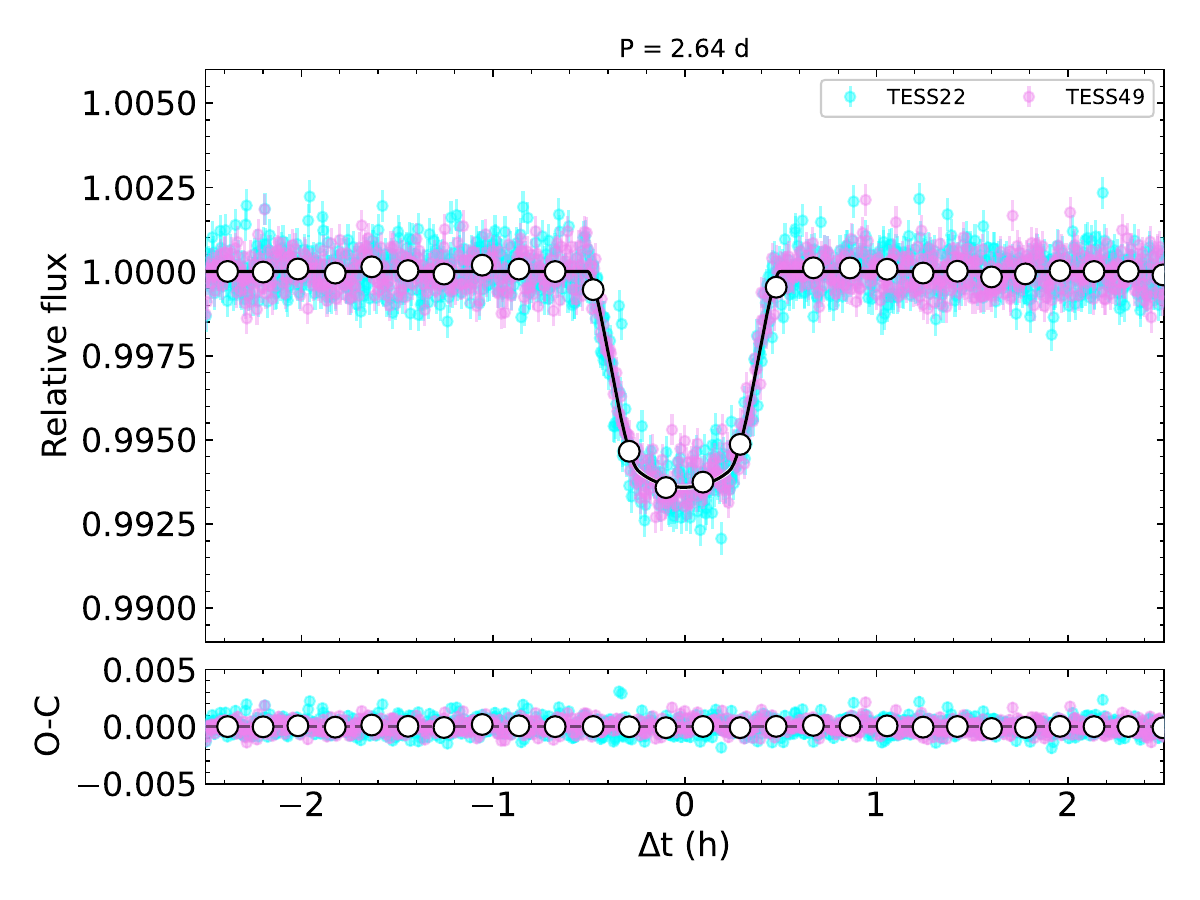}
        \caption{TESS light curves folded in phase with the orbital periods of GJ\,436\,b. Binned data are plotted as white circles. The best transit model is shown with a black line.
        }
        \label{fig:tesslc}
\end{figure}

\begin{figure}[h!]
        \centering
        \includegraphics[width=0.45\columnwidth]{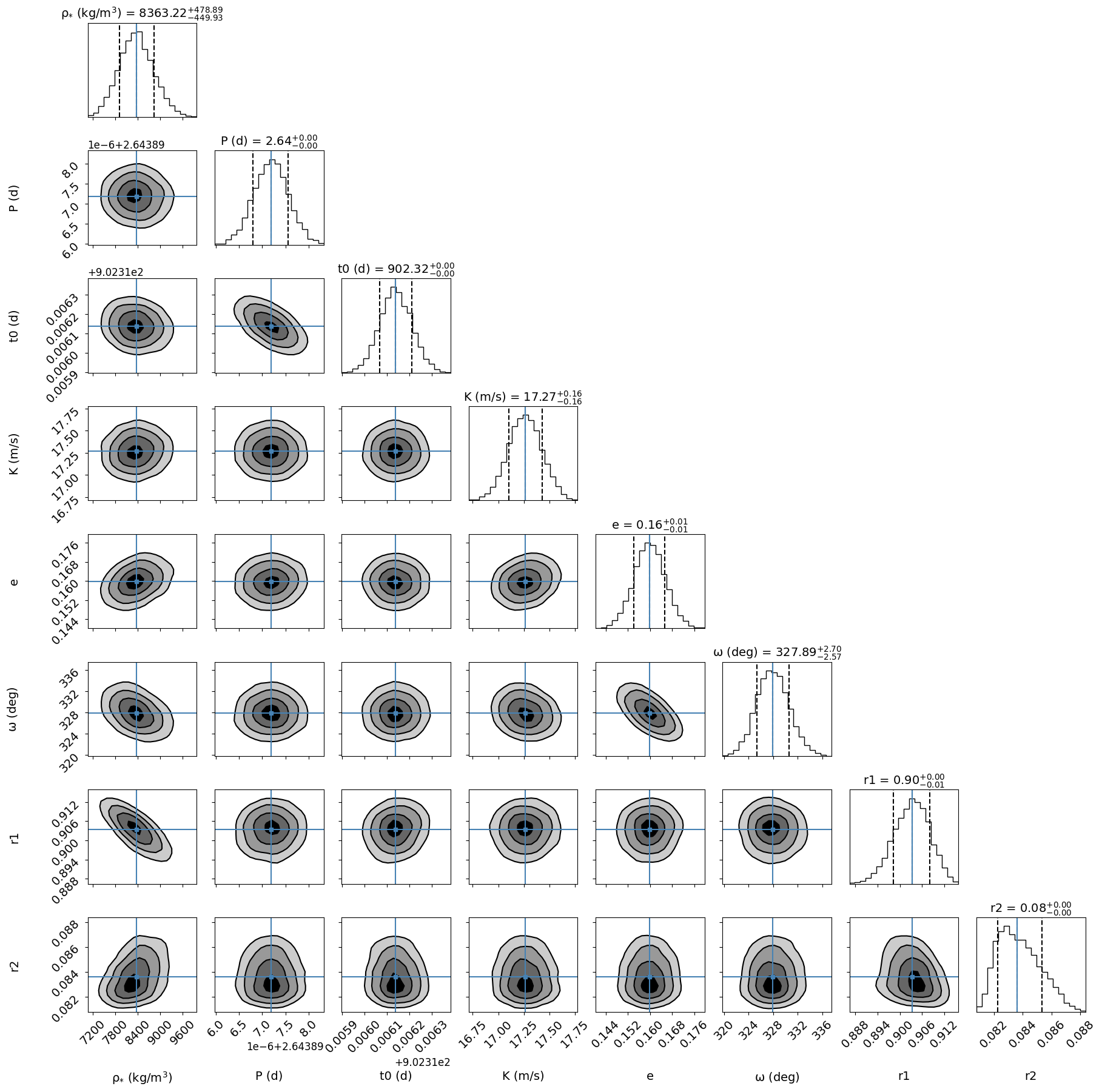}
        \caption{Posterior distributions of the principal fitted planetary parameters of the GJ\,436 system as obtained from the joint transit and RV fit, assuming eccentric orbit. The (black) vertical dashed lines indicate the 16\% and 84\%\,quantiles that were used to define the optimal values and their associated 1$\sigma$ uncertainty. The blue lines stands for the median  values (50\% quantile) of each fitted parameter.
        }
        \label{fig:gj436_cornerplot}
\end{figure}

\begin{figure}[h!]
        \centering
        \includegraphics[width=0.45\columnwidth]{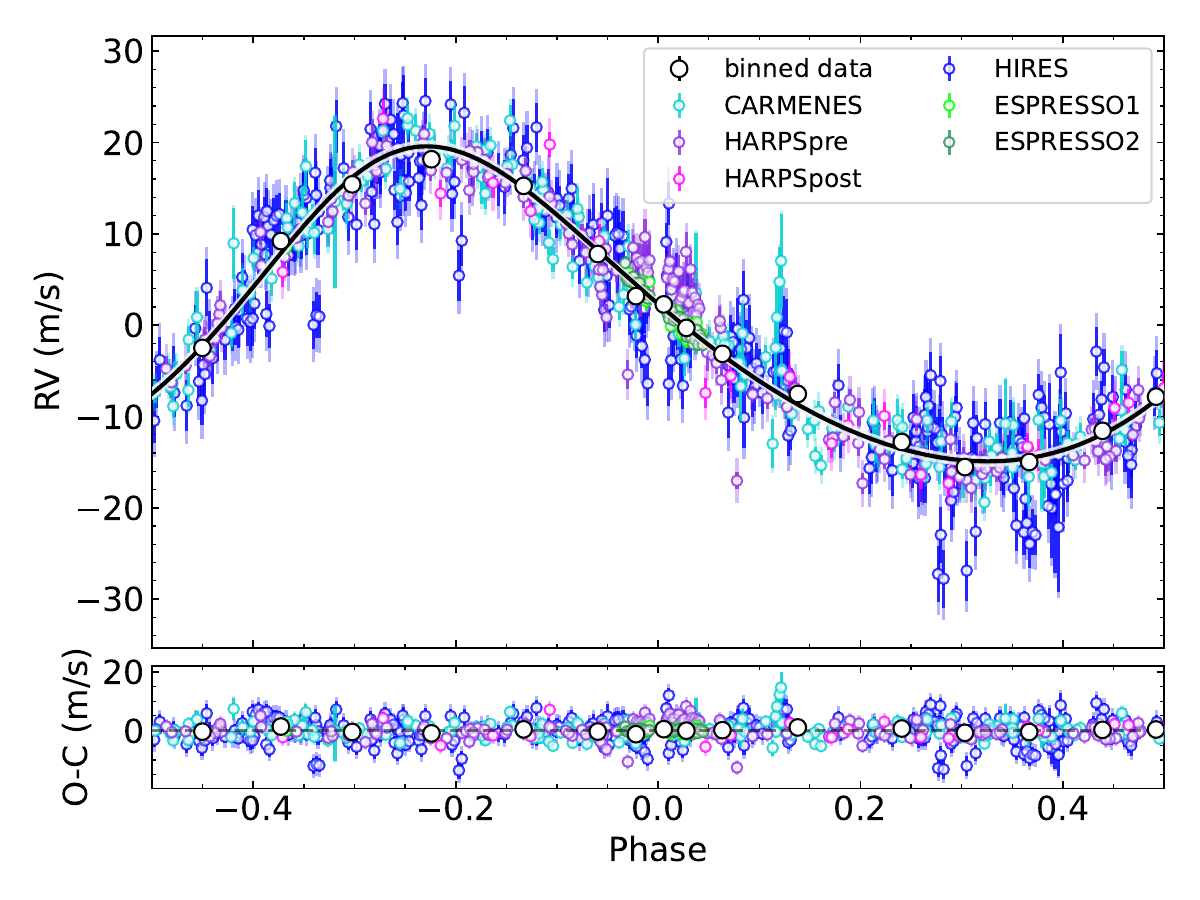}
        \caption{GJ\,436 HARPS, HIRES, ESPRESSO, and CARMENES RVs of GJ 436 (coloured dots) and the best model (black
                line) from the joint photometric and spectroscopic fit folded in phase with the orbital
                period of GJ\,436\,b, assuming eccentricity in the model.
        }
        \label{fig:gj436_RVmodel_vs_phase}
\end{figure}

\begin{figure}[h!]
    \centering
    \includegraphics[width=\columnwidth]{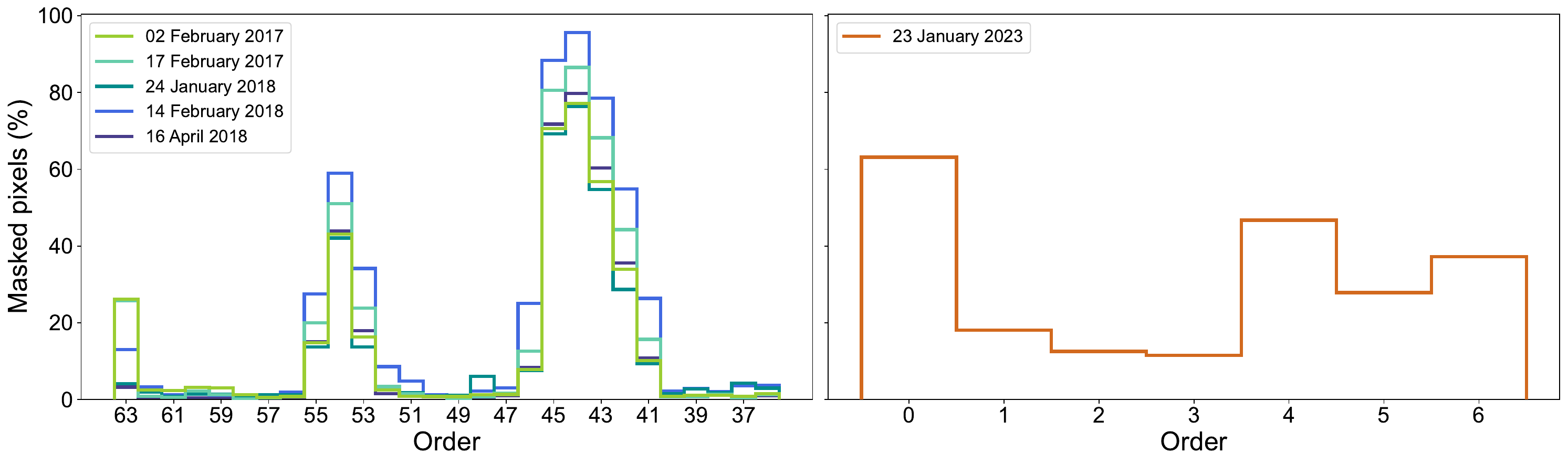}
    \caption{Percentage of masked pixels per spectral order for each observation night of the CARMENES (left) and CRIRES$^+$ (right) datasets. Higher percentages correspond to spectral orders containing the strongest telluric absorption bands, primarily due to water vapour.}
    \label{fig:masked_pix}
\end{figure}

\begin{figure}[h!]
    \centering
    \includegraphics[width=0.55\columnwidth]{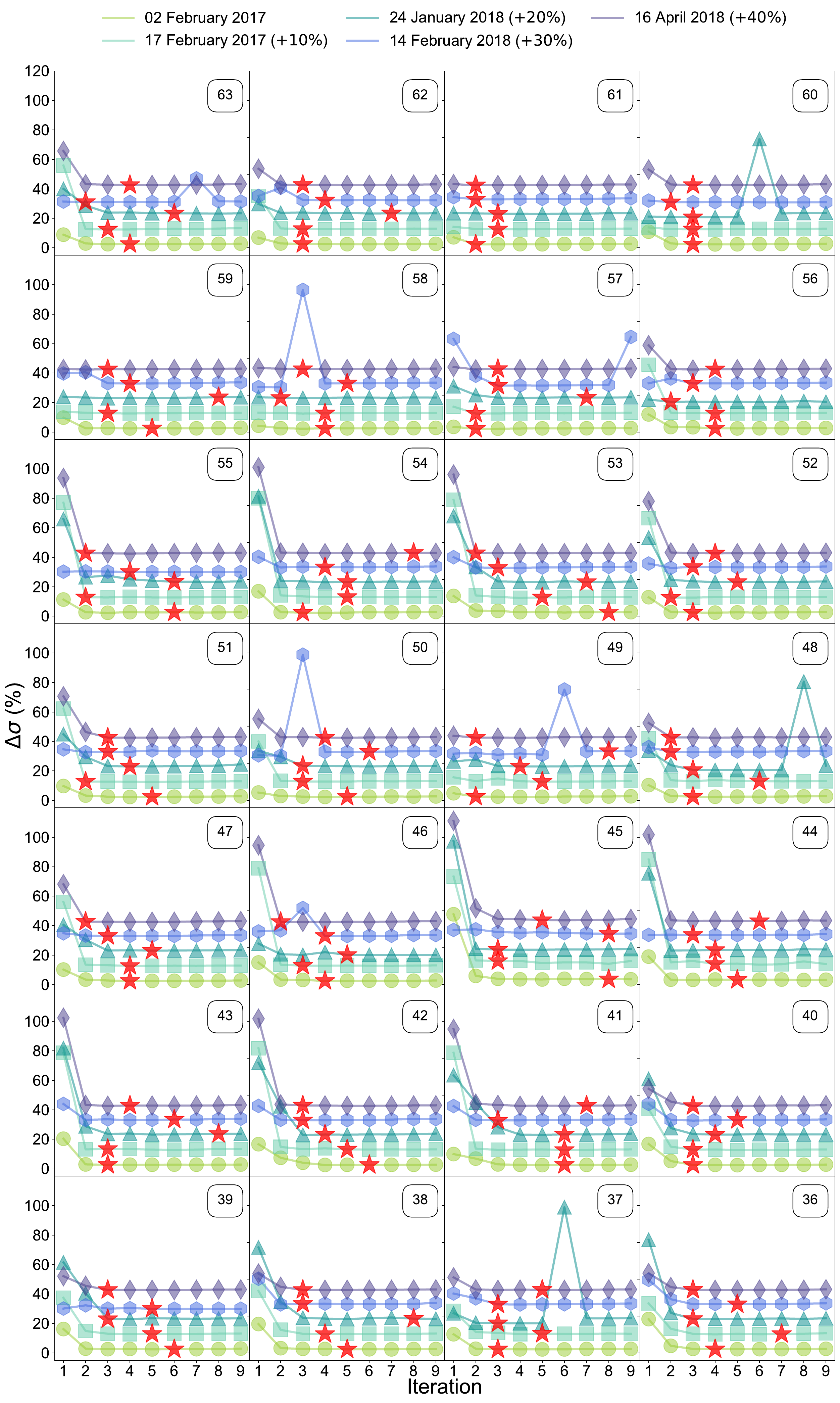}
    \caption{Illustration of the selection criterion used for order-wise {\tt SysRem} optimization for the CARMENES datasets. Here, the values of $\Delta\sigma$ (vertical axis) are plotted as a function of the {\tt SysRem} pass (horizontal axis) across all spectral orders for observations from all five nights. Red stars mark the {\tt SysRem} pass at which a plateau is reached, indicating the point at which the algorithm is halted. Spectral order number is indicated in the upper right corner of each cell. For clarity, an offset was added to the per-night curves, as indicated in the legend.}
    \label{fig:ds_per_order}
\end{figure}

\begin{figure}[h!]
    \centering
    \includegraphics[width=\textwidth]{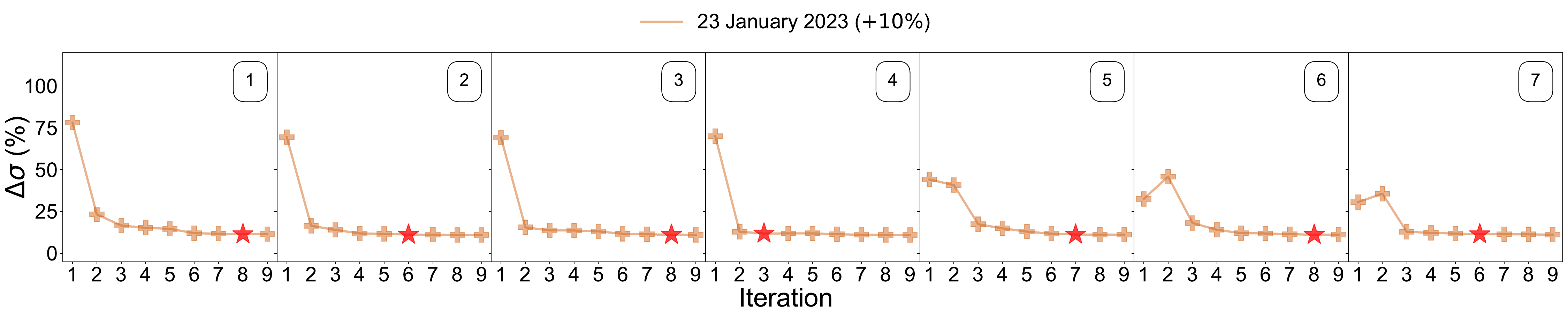}
    \caption{Same as Fig.\ref{fig:ds_per_order} but for the CRIRES$^+$ dataset.}
    \label{fig:ds_per_order_criresp}
\end{figure}

\begin{figure}[h!]
    \centering    \includegraphics[width=\textwidth]{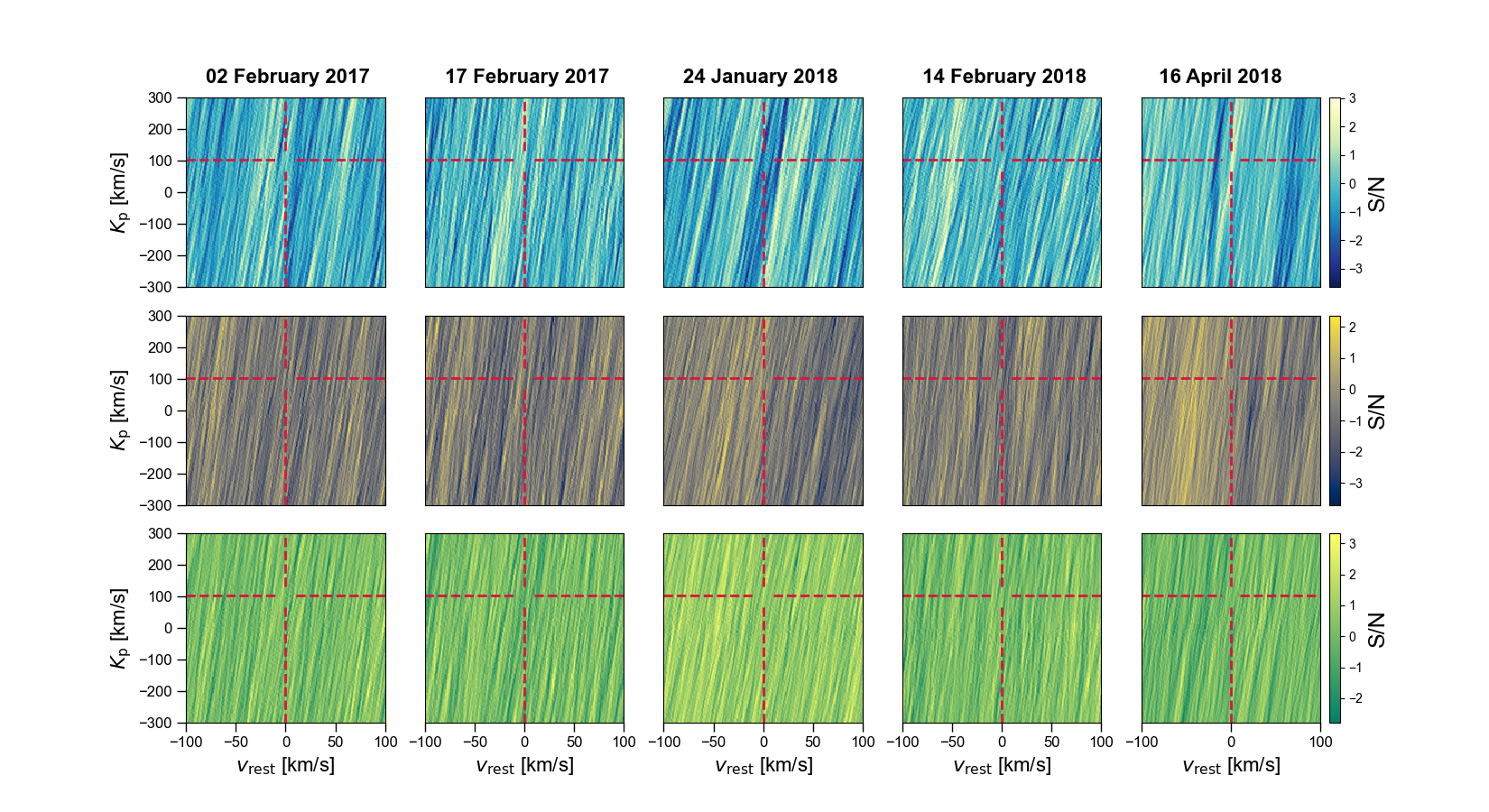}
    \caption{Same as Fig.\,\ref{fig:cc_water2} but for each individual CARMENES transit analysed and for the molecules studied. First row corresponds to water vapour, second row to methane, and third to carbon monoxide.}
    \label{fig:snr_individual}
\end{figure}

\begin{figure*}[h!]
    \centering
    \includegraphics[width=0.8\textwidth]{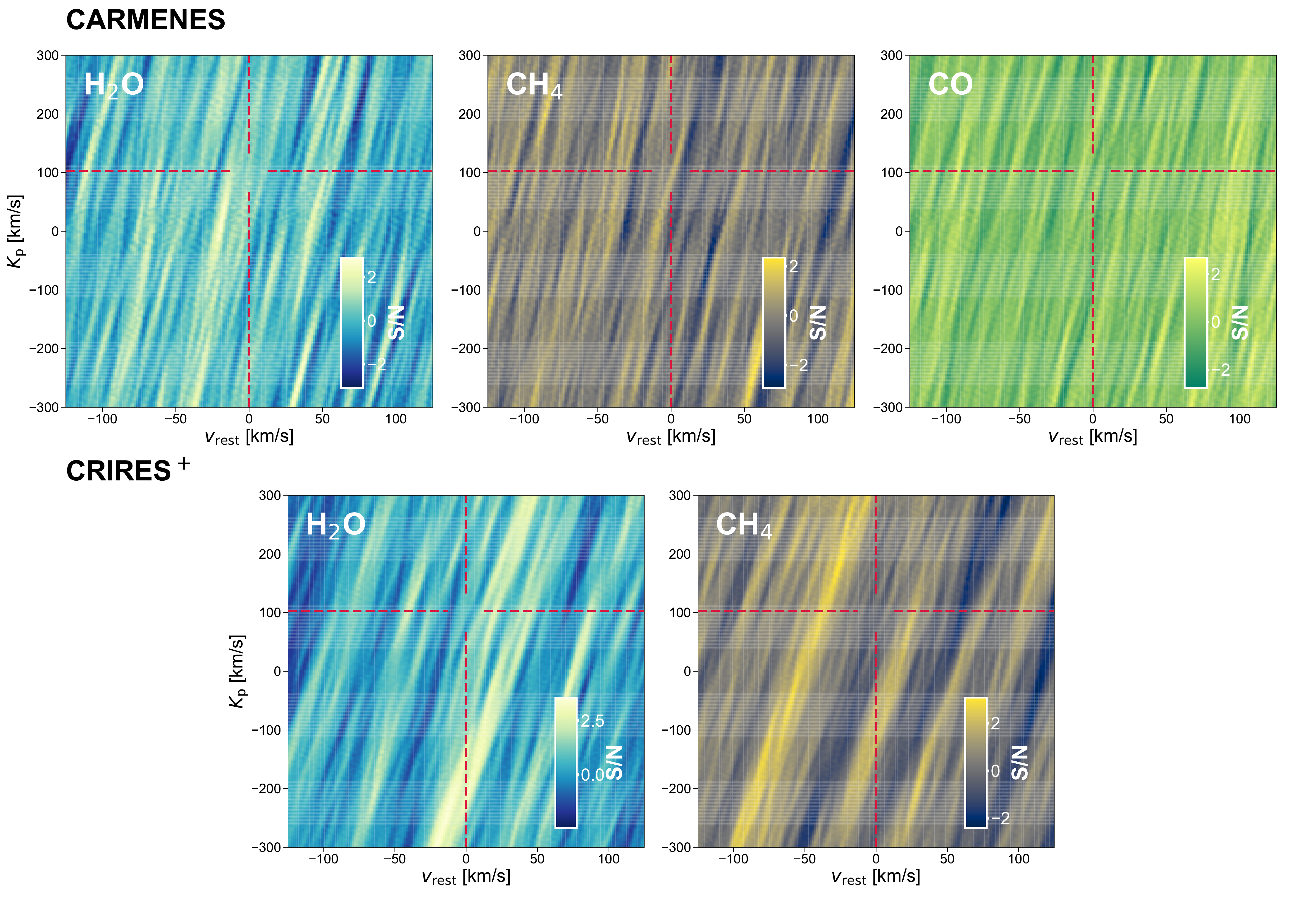}
    \caption{Cross-correlation maps in S/N units for potential atmospheric signals as a function of $v_{\rm rest}$ and $K_{\rm p}$ for the primary expected absorbing species in the atmosphere of GJ\,436\,b. For CARMENES (five combined transits; top row) and CRIRES$^+$ (one transit; bottom row) we explored, water vapour (left), methane (middle), and carbon monoxide (right, only for CARMENES). Horizontal and vertical red lines indicate the expected $K_{\rm p}$ and $v_{\rm rest}$ eccentric orbital models. These example maps were derived using a $10\times$\,solar metallicity template with a cloud deck at $10$\,mbar. No molecular signals were detected in our cross-correlation analyses.}
    \label{fig:all_mol_sum}
\end{figure*}

\begin{figure}[h!]
    \centering
    \includegraphics[width=230pt]{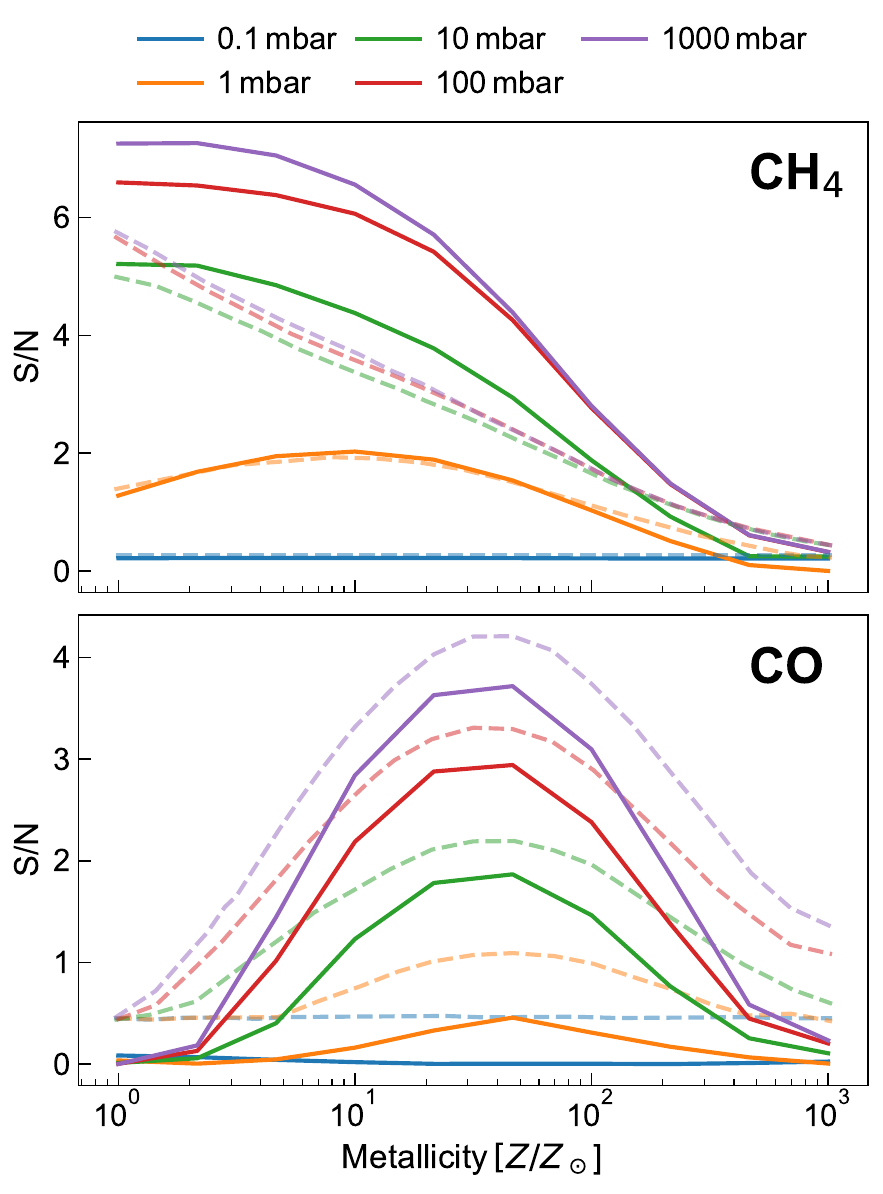}
    \caption{S/N slices from Fig.\,\ref{fig:injections} showing the S/N (vertical axis) as a function of the injected-model metallicity (horizontal axis). In the top panel, we show this result for CH$_4$, and CO is shown in the bottom panel. Different colours correspond to the different fixed $p{\rm_c}$. Dashed lines correspond to S/N slices at the same fixed $p{\rm_c}$ from Fig.\,$6$ of \cite{grasser2024}. Due to the unavailability of the original data, these values were extracted using \texttt{WebPlotDigitizer} (Version $5.2$).}
    \label{fig:slices_comparison}
    \end{figure}

\begin{figure}[h!]
\centering
\includegraphics[width=\textwidth]{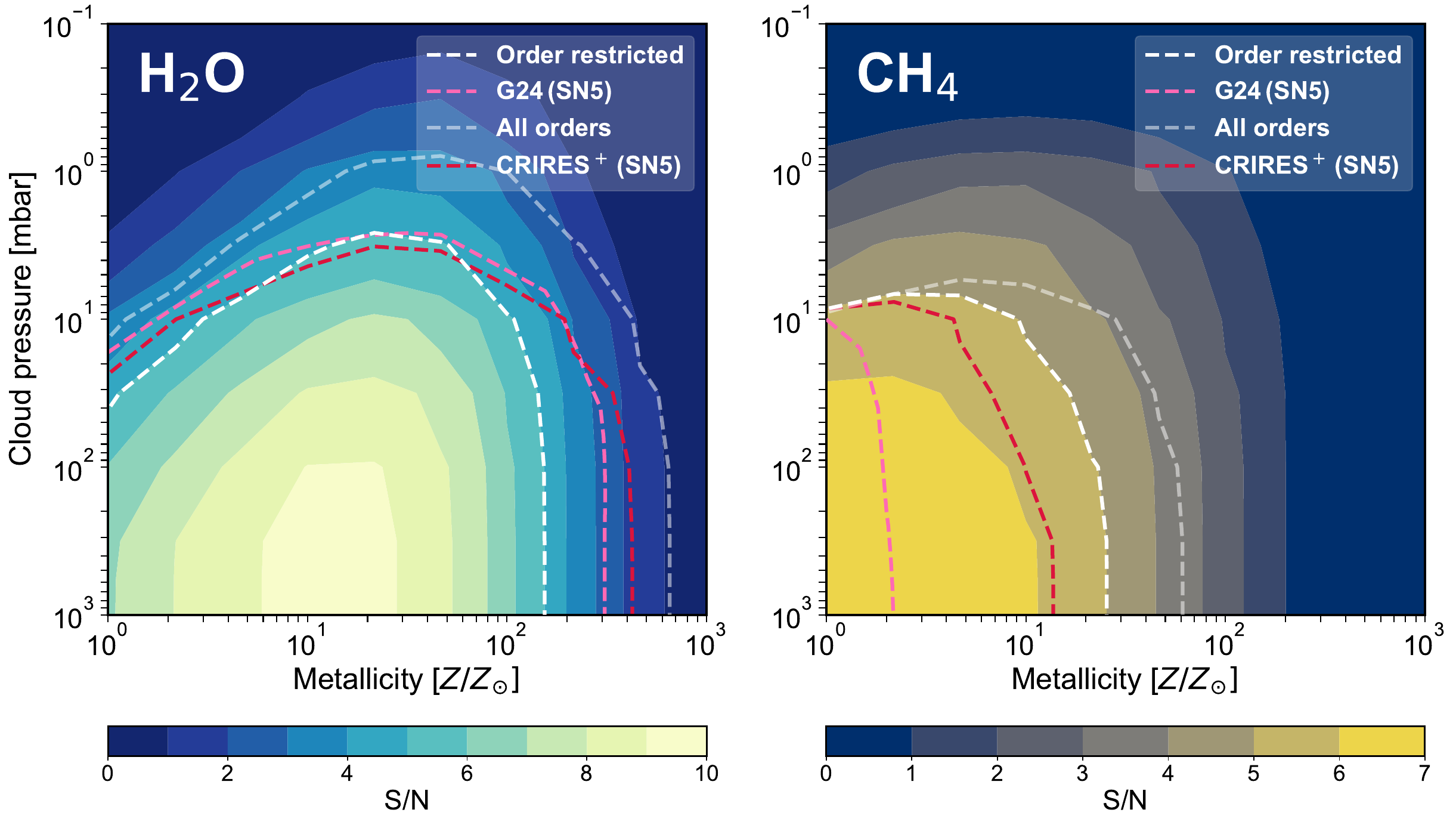}
\caption{Same as Fig.\,\ref{fig:injections} but for H$_2$O (left) and CH$_4$ (right) but limiting the CARMENES spectral range to closely match that of CRIRES$^+$.}
\label{fig:injections_7orders}
\end{figure}

\end{appendix}

\end{document}